\documentclass[sigconf]{acmart}

\usepackage{subcaption}
\usepackage{graphicx}
\usepackage{enumitem}
\usepackage{multirow}
\usepackage{soul}
\usepackage{stfloats}
\usepackage{makecell}
\usepackage[ruled,boxed,linesnumbered]{algorithm2e}
\usepackage{tabularx}
\newcounter{problemname}
\newcounter{definitionname}

\newenvironment{mydefinition}[1]{\par\vspace{0.3cm}\noindent\refstepcounter{definitionname}\textit{Definition \arabic{definitionname} ({#1}). }}{\par\vspace{0.3cm}}

\newcommand{\model}{{DoubleAdapt}~}
\newcommand{\modelns}{{DoubleAdapt}}
\newcommand{\eat}[1]{}
\AtBeginDocument{%
  }

\copyrightyear{2023}
\acmYear{2023}
\setcopyright{acmlicensed}
\acmConference[KDD '23] {Proceedings of the 29th ACM SIGKDD Conference on Knowledge Discovery and Data Mining}{August 6--10, 2023}{Long Beach, CA, USA.}
\acmBooktitle{Proceedings of the 29th ACM SIGKDD Conference on Knowledge Discovery and Data Mining (KDD '23), August 6--10, 2023, Long Beach, CA, USA}
\acmPrice{15.00}
\acmISBN{979-8-4007-0103-0/23/08}
\acmDOI{10.1145/3580305.3599315}

\graphicspath{ {./figures/} }

\begin{document}

\title{DoubleAdapt: A Meta-learning Approach to Incremental Learning for Stock Trend Forecasting}

\author{Lifan Zhao}
\affiliation{%
\institution{Shanghai Jiao Tong University}
\country{Shanghai, China}
}
\email{mogician233@sjtu.edu.cn}

\author{Shuming Kong}
\affiliation{%
\institution{Shanghai Jiao Tong University}
\country{Shanghai, China}}
\email{leinuo123@sjtu.edu.cn}

\author{Yanyan Shen}
\authornote{Yanyan Shen is the corresponding author.}
\affiliation{%
  \institution{Shanghai Jiao Tong University}
  \country{Shanghai, China}}
\email{shenyy@sjtu.edu.cn}

\renewcommand{\shortauthors}{Zhao et al.}

\begin{abstract}
Stock trend forecasting is a fundamental task of quantitative investment where precise predictions of price trends are indispensable. 
As an online service, stock data continuously arrive over time. 
It is practical and efficient to incrementally update the forecast model with the latest data which may reveal some new patterns recurring in the future stock market.
However, incremental learning for stock trend forecasting still remains under-explored due to the challenge of distribution shifts (\textit{a.k.a.} concept drifts). 
With the stock market dynamically evolving, the distribution of future data can slightly or significantly differ from incremental data, hindering the effectiveness of incremental updates.
To address this challenge, we propose \modelns, an end-to-end framework with two adapters, which can effectively adapt the data and the model to mitigate the effects of distribution shifts. 
Our key insight is to automatically learn how to adapt stock data into a locally stationary distribution in favor of profitable updates. Complemented by data adaptation, we can confidently adapt the model parameters under mitigated distribution shifts. We cast each incremental learning task as a meta-learning task and automatically optimize the adapters for desirable data adaptation and parameter initialization. Experiments on real-world stock datasets demonstrate that \model achieves state-of-the-art predictive performance and shows considerable efficiency. 
Our code is available at {https://github.com/SJTU-Quant/qlib/}.
\end{abstract}




\maketitle

\section{Introduction}
Stock trend forecasting, which aims at predicting future trends of stock prices, is a fundamental task of quantitative investment and has attracted soaring attention in recent years~\cite{TRA, REST}. Due to the widespread success of deep learning, various neural networks have been developed to exploit intricate patterns of the stock market and infer future price trends. As an online application, new stock data arrive in a streaming way as time goes by. This gives rise to an increasingly growing dataset that is enriched with more underlying patterns. 
It is of vital importance to continually learn new emerging patterns from incoming stock data, in order to avoid the model aging issue~\cite{LLF} and pursue higher accuracy in future predictions.

To this end, a common practice named \textit{Rolling Retraining} (RR) is used to periodically leverage the whole enlarged dataset to retrain the model parameters from scratch. However, RR usually leaves out abundant recent samples for validation and fails to retrain the model on the validation set, of which the patterns are often informative and valuable for future predictions~\cite{DDGDA}. 
Another fatal drawback of RR lies in its expensive time and space consumptions. The training time increases with the size of the enlarged training data, which further causes an unbearable duration of hyperparameter tuning and retraining algorithm selection. 
An alternative way known as \textit{Incremental Learning} (IL) is to fine-tune the model \textit{only} with the latest incremental data. In each IL task, the model is initialized by inheriting parameters from the preceding IL task that are expected to memorize historical patterns, and then the model is consolidated with new knowledge in incremental data. The rationale behind it is that recent data may reveal some new patterns that did not appear before but will reoccur in the future. 
Moreover, IL is not only dramatically faster but also occupies much smaller space than RR. 

\begin{figure}[t]
  \vspace{4pt}
	\centering
  \captionsetup[subfigure]{skip=1pt}
	\subcaptionbox{Gradual shift\label{fig:shift:a}}
	{
    \includegraphics[width=.98\linewidth]{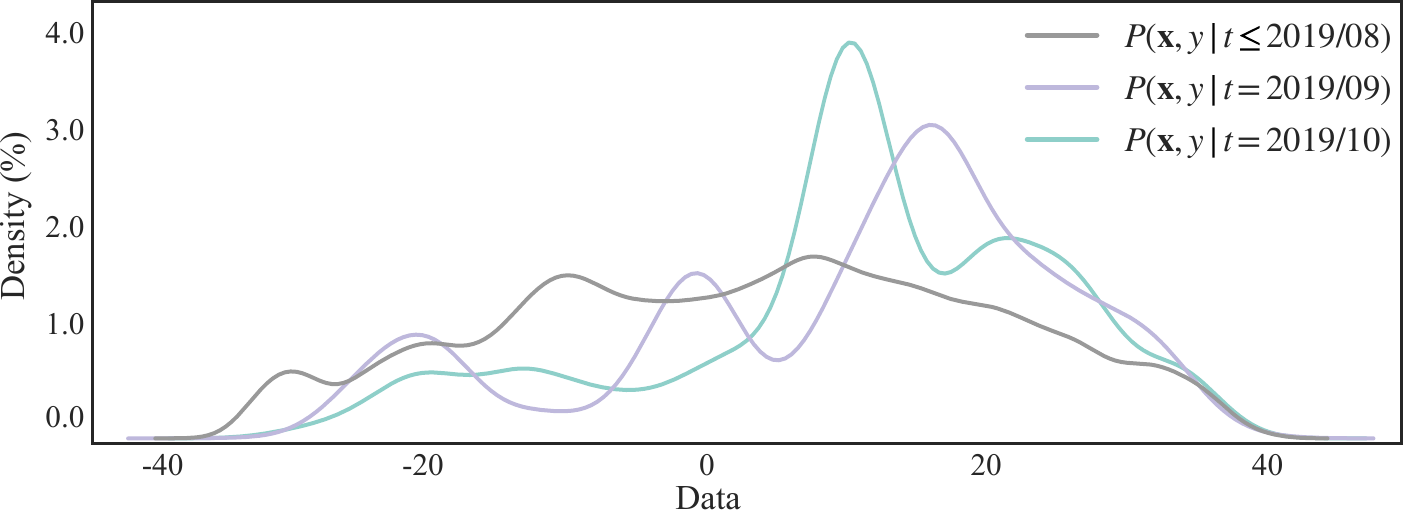}
  }\\
  \vspace{4pt}
	\subcaptionbox{Abrupt shift\label{fig:shift:b}}
	{
    \includegraphics[width=0.98\linewidth]{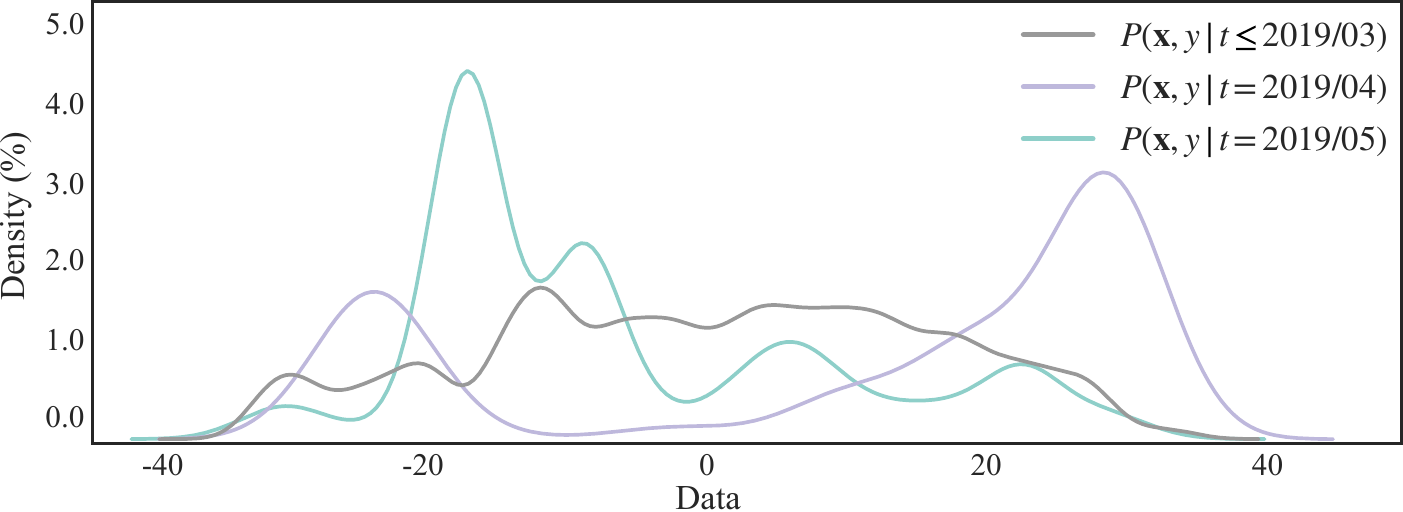}
  }\\
	\caption{{Illustration of distribution shifts in CSI 300 stock set. The vector of each stock including dozens of technical indicators and the corresponding label is mapped via t-SNE~\cite{t-SNE} to a 1-D point on the horizontal axis. We estimate the distributions with a kernel density estimator. We plot the distribution of the incremental data in one month in purple, the distribution of all the previous data in black, and the distribution of the next month's data in green.}}
  \label{fig:shift}
\end{figure}

Despite its considerable efficiency and potential effectiveness, IL is still under-explored in stock trend forecasting mainly due to the challenge of \textbf{distribution shifts} (\textit{a.k.a.} concept drifts).
IL performs well only if there is always little difference between the distributions of incremental data and future data.
However, it is well accepted that the stock market is in a non-stationary environment where data distribution irregularly shifts over time~\cite{DDGDA, LLF, MASSER}. Such distribution shifts can vary in direction and degree. For example, in Figure~\ref{fig:shift}, we visualize two real cases of distribution shifts in the Chinese stock market. Given historical samples in the last month (\textit{e.g.}, 2019/09) as \textit{incremental data}, we are meant to deploy a model online and do inference on the test samples in the next month (\textit{e.g.}, 2019/10), termed as \textit{test data}.
In the case of gradual shifts as shown in Figure~\ref{fig:shift:a}, incremental data can reveal some future tendencies but its distribution still differs from the test data. 
A model that well fits the incremental data may not perfectly succeed in the future. Moreover, incremental data could even become misleading when distribution shifts abruptly appear, making a nonnegligible gap between the two data distributions, as shown in Figure~\ref{fig:shift:b}. 
As long as distribution shifts exist, typical IL cannot consistently benefit from incremental data and may even suffer from inappropriate updates.
In a nutshell, the discrepancy between the distributions of incremental data and test data could hinder the overall performance, posing the key challenge to IL for stock trend forecasting.

Confronted with this challenge, it is noteworthy that the incremental updates stem from two factors: the incremental data and the initial parameters. Conventional IL blindly inherits the parameters learned in the previous task as initial parameter weights and conducts one-sided model adaptation on raw incremental data. To improve IL against distribution shifts, we propose to strengthen the learning scheme by performing two-fold adaptation, namely \textbf{data adaptation} and \textbf{model adaptation}. The data adaptation aims to close the gap between the distributions of incremental data and test data. For example, biased patterns that only exist in incremental data are equivalent to noise with respect to test data and could be resolved through proper data adaptation. Our model adaptation focuses on learning a good initialization of parameters for each IL task, which can appropriately adapt to incremental data and still retain a degree of robustness to distribution shifts. However, it is intractable to design optimal adaptation for each IL task. A proper choice of adaptation varies by forecast model, dataset, period, degree of distribution shifts, and so on. Hence, we borrow ideas from meta-learning~\cite{MAML} to realize the two-fold adaptation, \textit{i.e.}, to automatically find profitable data adaptation without human labor or expertise and to reach a sweet spot between adaptiveness and robustness for model adaptation.

In this work, we propose \modelns, a meta-learning approach to incremental learning for stock trend forecasting. 
We introduce two meta-learners, namely data adapter and model adapter, which adapt data towards a locally stationary distribution and equip the model with task-specific parameters that have quickly adapted to the incremental data and still generalize well on the test data. 
The data adapter contains a multi-head feature adaptation layer and a multi-head label adaptation layer in order to obtain adapted incremental data and adapted test data that are profitable for incremental learning. Specifically, the feature adaptation layer transforms all features from the incremental data and the test data, while the label adaptation layer rectifies labels of the incremental data and its inverse function restores model predictions on the test data. By casting the problem of IL for stock trend forecasting as a sequence of meta-learning tasks, we perform each IL task by solving a bi-level optimization problem: 
(i) in the lower-level optimization, the parameters of the forecast model are initialized by the model adapter and fine-tuned on the adapted incremental data; 
(ii) in the upper-level optimization, the meta-learners are optimized by test errors 
on the adapted test data.
Throughout the online inference phase, both the two adapters and the forecast model will be updated continually over new IL tasks. 

The main contributions of this work are summarized as follows.

\begin{itemize}[leftmargin=*]
  \item We propose \modelns, an end-to-end incremental learning framework for stock trend forecasting, which adapts both the data and the model to cope with distribution shifts in the online environment.   
  \item We formulate each incremental learning task as a bi-level optimization problem. The lower level is the forecast model that is initialized by the model adapter and fine-tuned using the adapted incremental data. 
  The upper level includes the data adapter and the model adapter as two meta-learners that are optimized to minimize the forecast error on the adapted test data. 
  \item We conduct experiments on real-world datasets and demonstrate that \model performs effectively against different kinds of distribution shifts and achieves state-of-the-art predictive performance compared with RR and meta-learning methods. \model also enjoys high efficiency compared with RR methods.
\end{itemize}

\section{Preliminaries}

In this section, we will introduce some definitions of our work and formulate the incremental learning problem. We also highlight the challenge of distribution shifts.

\begin{mydefinition}{Stock Price Trend}\label{def:trend}
  Following~\cite{HAN2018, REST, HIST}, we define the stock price trend at date $t$ as the stock price change rate of the next day:
  \begin{equation}
    y^{(t)} = \frac{Price^{(t+1)} - Price^{(t)}}{Price^{(t)}},
  \end{equation}
  where $Price^{(t)}$ is the closing price at date $t$ and could also be the opening price or volume-weighted average price (VWAP).
\end{mydefinition}

Let $\mathbf x^{(t)} \in \mathbb R^{D}$ represent the feature vector of a stock at date $t$, where $D$ is the feature dimension. For example, we can constitute $\mathbf x$ with opening price, closing price, and other indicators in recent days. Its stock price trend $y$ is the corresponding label. Suppose the stock market comprises $S$ stocks. The collection of features and labels of the $S$ stocks at date $t$ can be denoted as $\mathbf X^{(t)} \in \mathbb R^{S\times D}$ and $\mathbf Y^{(t)} \in \mathbb R^{S}$, respectively.
The goal of stock trend forecasting is to learn a forecast model $F$ on historical data $\{(\mathbf{X}^{(t)}, \mathbf Y^{(t)})\}_{t=1}^T$ and then forecast the labels of future data $\{(\mathbf{X}^{(t)}, \mathbf Y^{(t)})\}_{t=T+1}^{T'}$, where $T$ and $T'$ are the end time of the historical data and the future data, respectively. The parameters of $F$ are denoted as $\theta$. 

\begin{figure}[t]
  \centering
  \includegraphics[width=\linewidth]{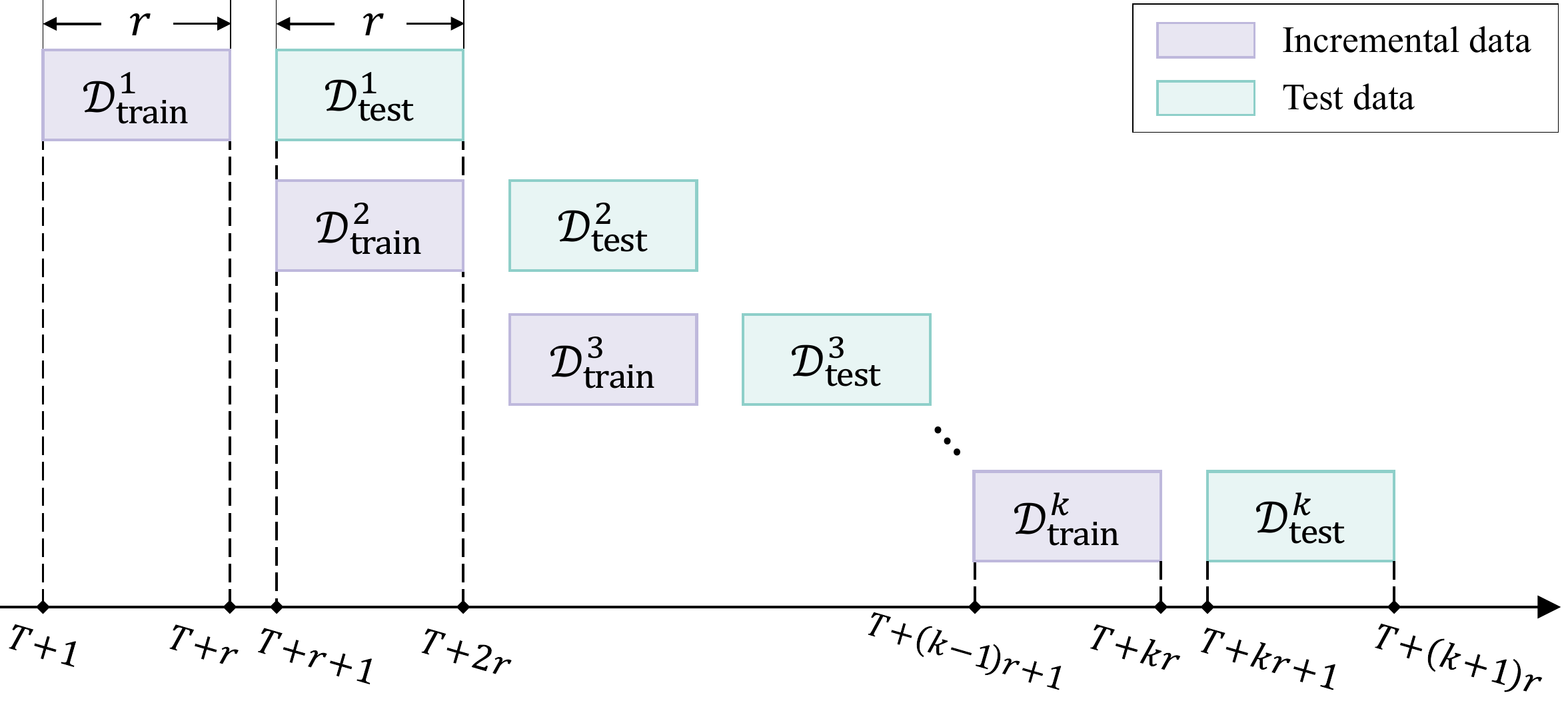}
  \caption{Illustration of IL for stock trend forecasting. $r$ is the timespan of incremental data or test data. 
  }
  \label{fig:IL}
\end{figure}

In online scenarios, new data are coming over time. Once the ground-truth labels of new samples are obtained, we can update the forecast model to learn new emerging patterns. 
In this work, we focus on incremental learning (IL) for stock trend forecasting, where we periodically launch IL tasks to update the model \textit{only} with incremental data, as illustrated in Figure~\ref{fig:IL} and defined as follows.

\begin{mydefinition}{IL Task for Stock Trend Forecasting}
  Supposing a pretrained forecast model is deployed online at date $T$+$1$, we launch an IL task every $r$ dates, where $r$ is predetermined by practical applications. For the $k$-th IL task at date $T$+$kr$+$1$, we fine-tune the model parameters $\theta^{k-1}$ on incremental data ${\mathcal D}^{k}_{\text{train}}$ and predict labels on test data ${\mathcal D}^{k}_{\text{test}}$ in the following $r$ dates, where ${\mathcal D}_{\text{train}}^{k}=\{(\mathbf{X}^{(t)}, \mathbf Y^{(t)})\}_{t=T+(k-1)r+1}^{T+kr}$ and ${\mathcal D}^{k}_{\text{test}}=\{(\mathbf{X}^{(t)}, \mathbf Y^{(t)})\}_{t=T+kr+1}^{T+(k+1)r}$. 
  The outputs of the $k$-th task are updated parameters $\theta^{k}$ and predictions $\{ \hat{\mathbf Y}^{(t)}\}_{i=T+kr+1}^{T+(k+1)r}$. 
  The IL task expects the model to quickly adapt to the incremental data, in order to make precise predictions on future data with similar patterns. We can evaluate the predictions by computing a loss function $\mathcal L_{\text{test}}$ on ${\mathcal D}^{k}_{\text{test}}$, \textit{e.g.}, mean square error.
\end{mydefinition}

\noindent\textbf{Problem Statement. }
  Given a predefined task interval $r$, IL for stock trend forecasting is constituted by a sequence of IL tasks, \textit{i.e.}, $\mathcal{T} = $
  $\{({\mathcal D}_{\text{train}}^{1}, {\mathcal D}_{\text{test}}^{1})$, 
  $\ ({\mathcal D}_{\text{train}}^{2}, {\mathcal D}_{\text{test}}^{2})$,
  $\ \cdots, ({\mathcal D}_{\text{train}}^{k}, {\mathcal D}_{\text{test}}^{k}), \cdots\}$. 
  In each task, we update the model parameters, do online inference, and end up with performance evaluation on the ground-truth labels of the test data. The goal of IL is to achieve the best overall performance across all test dates, which can be evaluated by excess annualized returns or other ranking metrics of stock trend forecasting. 

\vspace{0.2em}
Typically, IL holds a strong assumption that a model which fits recent data can perform well on the following data under the same distribution. However, as the stock market is dynamically evolving, its data distribution can easily shift over time.
$\mathcal D_{\text{train}}^{k}$ and $\mathcal D_{\text{test}}^{k}$ are likely to have two different joint distributions, \textit{i.e.}, $\mathcal P_{\text{train}}^{k}(\mathbf x, y) \ne \mathcal P_{\text{test}}^{k}(\mathbf x, y)$, where $\mathcal P_{\text{train}}^{k}$ and $\mathcal P_{\text{test}}^{k}$ denote the distributions of $\mathcal D_{\text{train}}^{k}$ and $\mathcal D_{\text{test}}^{k}$, respectively. 
The distribution shifts can be zoomed into the following two cases~\cite{SurveyConcepDrift}: 
\begin{itemize}[leftmargin=*]
  \item \textit{Conditional distribution shift} when $\mathcal P_{\text{train}}^{k}(y\mid\mathbf x) \ne \mathcal P_{\text{test}}^{k}(y\mid\mathbf x)$;
  \item \textit{Covariate shift} when $\mathcal P_{\text{train}}^{k}(\mathbf x) \ne \mathcal P_{\text{test}}^{k}(\mathbf x)$ and $\mathcal P_{\text{train}}^{k}(y\mid\mathbf x) = \mathcal P_{\text{test}}^{k}(y\mid\mathbf x)$.
\end{itemize}
In either case, excessive updates on incremental data would incur the overfitting issue while deficient updates may result in an underfit model. 
Hence, distribution shifts pose challenges to incremental learning for stock trend forecasting.

\section{Key Insights}

To tackle the distribution shift challenge, there are two important directions to follow. One is to close the gap between incremental data and test data so as to perform IL on more stationary distributions. Another is to enhance the generalization ability of the model against distribution shifts. Following both directions, we propose to perform data adaptation and model adaptation for each IL task. We describe our key insights with the following details.



\vspace{0.2cm}
\noindent\textbf{Data Adaptation.} A critical yet under-explored direction is to adapt data into a locally stationary distribution so as to mitigate the effects of distribution shifts at the data level. Some RR methods resample all historical data (\textit{e.g.}, 500 million samples) into a new training set that shares a similar distribution with future data~\cite{DDGDA}. However, such a \textit{coarse-grained} adaptation fails in IL where incremental data is of limited size (\textit{e.g.}, one thousand samples) and contains deficient samples to reveal future patterns. To address this limitation, we propose to adapt all features and labels of the incremental data to mitigate the effects of distribution shifts in a \textit{fine-grained} way. 
We argue that some shift patterns repeatedly appear in the historical data and are learnable. 
For example, stock prices can overreact to some bullish news and emotional investment, while the prices and the trend patterns tend to shift towards normal in future weeks.
Thus, it is often desirable to retract the overreacting features and labels to approach future tendencies. 
In addition, assuming the original incremental data is reliable, test data of a different distribution can be deemed as a biased dataset. 
Adapting test data towards the distribution of incremental data has a debiasing effect. 
Hence, we adapt both $\mathcal {D}_{\text {train}}^{k}$ and $\mathcal {D}_{\text {test}}^{k}$ so as to narrow the gap between their distributions.


\begin{figure}[ht]

  \centering
  \setlength{\abovecaptionskip}{6pt}
  \setlength{\belowcaptionskip}{-15pt}
  \includegraphics[width=\linewidth]{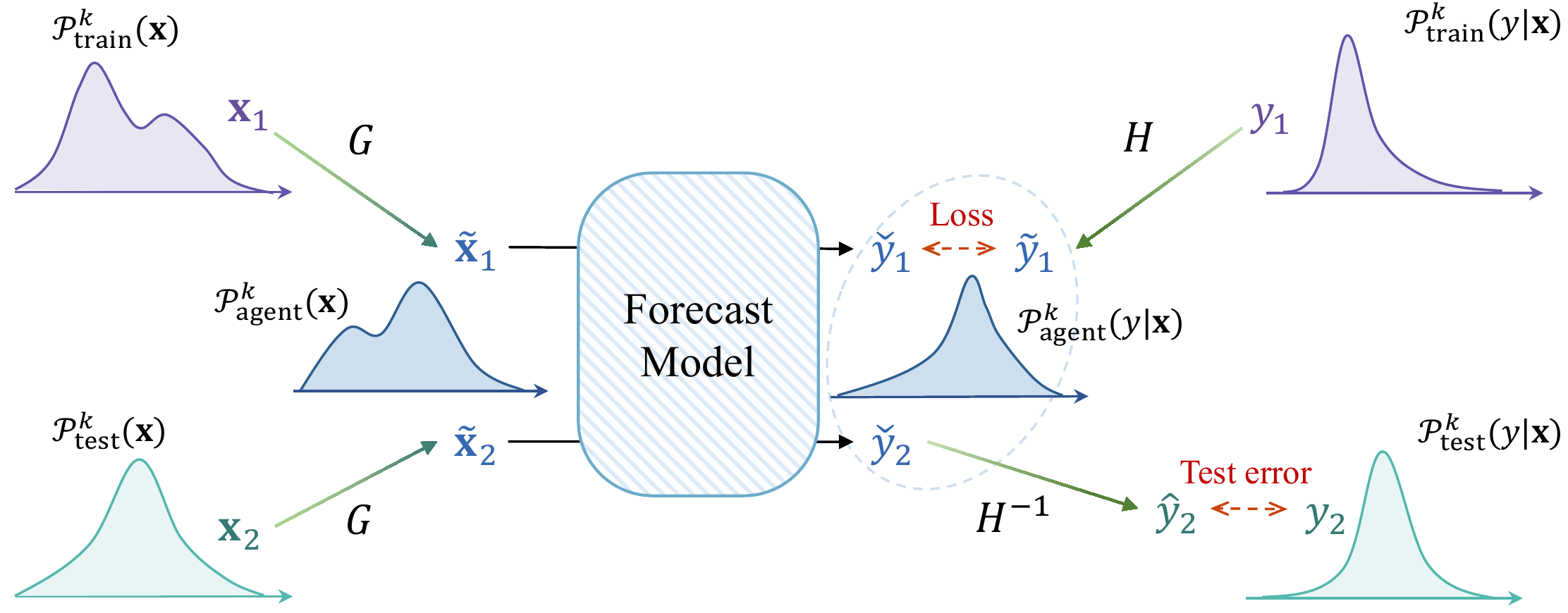}
  \caption{Illustration of data adaptation. The function $G$ adapts the feature distributions of the incremental data and the test data to an agent feature distribution. Similarly, the function $H$ adapts the two posterior distributions to an agent one. Its inverse function $H^{-1}$ restores the model outputs to the original posterior distribution of the test data.}
  \label{fig:flow}
\end{figure}

Technically, we cannot directly align $\mathcal P_{\text{train}}^{k}(\mathbf x, y)$ and $\mathcal P_{\text{test}}^{k}(\mathbf x, y)$ because labels of test data are unknown at the inference time. We thus decouple distribution shifts into covariate shifts and conditional distribution shifts, and address them separately, as illustrated in Figure~\ref{fig:flow}. 
\textbf{First}, we require a mapping function $G$ to transform the features of $\mathcal {D}_{\text {train}}^{k}$ and $\mathcal {D}_{\text {test}}^{k}$ in a fine-grained way, and we expect $G$ to adapt $\mathcal P_{\text{train}}^{k}(\mathbf x)$ and $\mathcal P_{\text{test}}^{k}(\mathbf x)$ to an agent feature distribution $\mathcal P_{\text{agent}}^{k}(\mathbf x)$, alleviating covariate shifts. 
\textbf{Second}, we apply another mapping function $H$ to adapt the labels of $\mathcal {D}_{\text {train}}^{k}$. We expect $H$ to adapt $\mathcal {P}_{\text{train}}^{k}(y|\mathbf x)$ to a possible future distribution $\mathcal P_{\text{agent}}^{k}(y|\mathbf x)$, dealing with conditional distribution shifts. 
Ideally, $H$ could project the labels from $\mathcal P_{\text{test}}^{k}(y|\mathbf x)$ into $\mathcal P_{\text{agent}}^{k}(y|\mathbf x)$ so that a forecast model fitting $\mathcal P_{\text{agent}}^{k}(y|\mathbf x)$ can precisely predict the adapted label for the test data. Finally, we need to inversely map the model outputs from $\mathcal P_{\text{agent}}^{k}(y| \mathbf x)$ to $\mathcal P_{\text{test}}^{k}(y| \mathbf x)$.
\textbf{As such}, we narrow the gap between $\mathcal P_{\text{train}}^{k}(\mathbf x)$ and $\mathcal P_{\text{test}}^{k}(\mathbf x)$ via feature adaptation, and narrow the gap between $\mathcal {P}_{\text{train}}^{k}(y|\mathbf x)$ and $\mathcal {P}_{\text{test}}^{k}(y|\mathbf x)$ via label adaptation. This allows us to reduce the discrepancy between the joint distributions, alleviating the distribution shift issue.

 \vspace{0.2cm}
\noindent\textbf{Model Adaptation.}  
Typically, IL initializes the model by the parameters learned in the previous task and updates the initial parameters on incremental data. Distribution shifts would hinder the test performance if the parameters after updates fall into a local optimum and overfit the incremental data. This motivates us to learn a good initialization of parameters for each IL task. 
On the one hand, the initial parameters of each task is required to preserve historical experience and retain generalization ability against distribution shifts. On the other hand, the parameters in IL still need to effectively memorize task-specific information without being trapped in past experiences. Therefore, we emphasize another important direction where we optimize the initial parameters of each IL task for robustness and adaptiveness.


 \vspace{0.2cm}
\noindent\textbf{Optimization via Meta-learning.} Following our key insights, we aim to mitigate distribution shifts at the data level and enhance generalization ability at the parameter level.
Desirable data adaptation should make the distributions of the two datasets more similar but still informative for stock trend forecasting.
Nevertheless, it is infeasible to manually design proper data adaptation as numerous factors should be considered, \textit{e.g.}, forecast models, datasets, prediction time, degrees of distribution shifts, and so on. 
For model adaptation, it is also tough to reach a sweet spot between robustness and adaptiveness. We thus introduce a meta-learning optimization objective to guide profitable data adaptation and parameter initialization. 
Though some normalization techniques ~\cite{MetaNorm, InstanceNorm, Tasknorm} in meta-learning can reduce distribution divergences, they may destroy the original statistical indicators (\textit{i.e.}, mean and standard deviation), which are critical task-specific information and deserve memorization in the online settings. In light of this, we pioneer mitigating the distribution shifts in meta-learning through neural networks rather than normalization.


\section{Methodology}


\subsection{Overview}

\begin{figure}[t]
  \setlength{\belowcaptionskip}{-8pt}
  \centering
  \includegraphics[width=\linewidth]{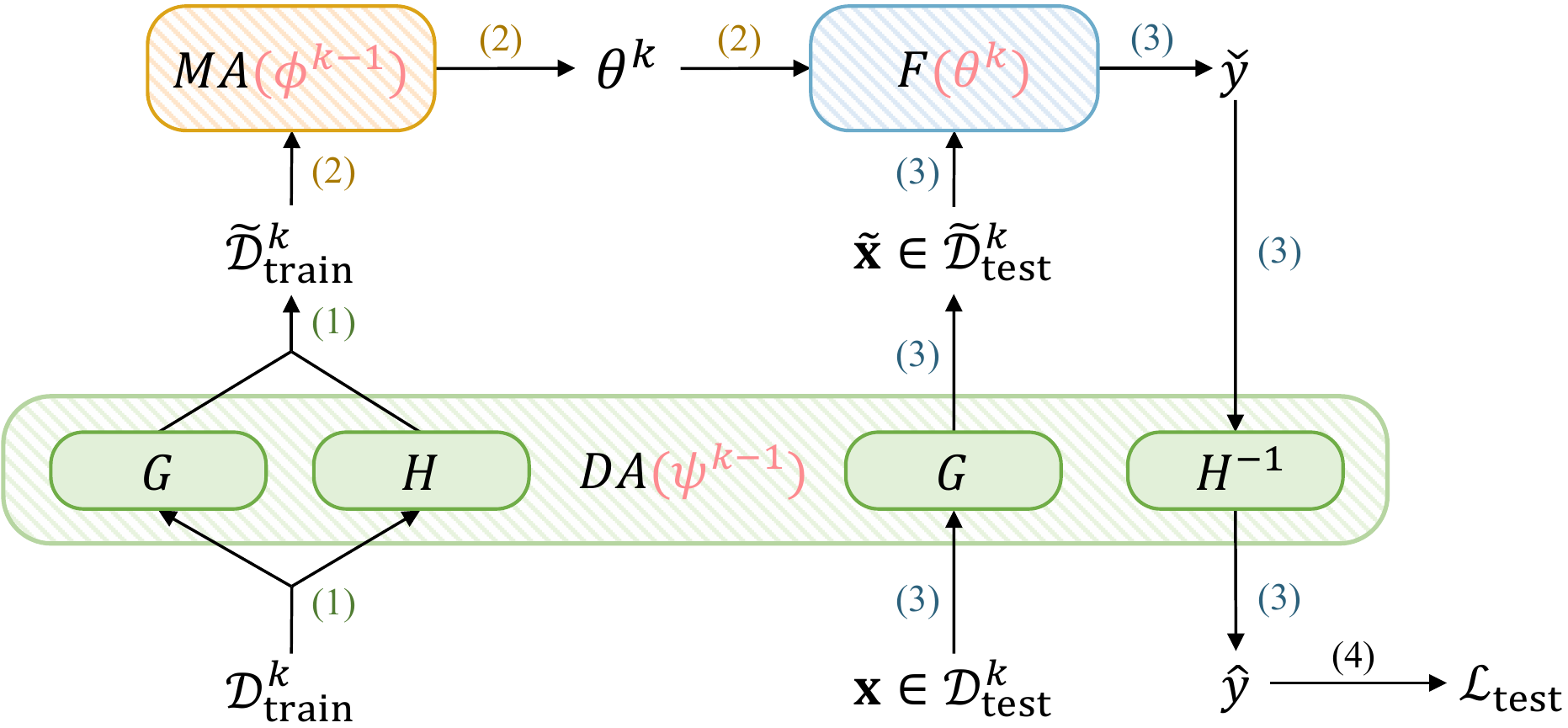}
  \caption{Overview of \model with a data adapter ${DA}$ and a model adapter ${MA}$. The parameters are shown in red.}
  \label{fig:overview}
\end{figure}

Figure~\ref{fig:overview} depicts the overview of our \model framework, which consists of three key components: \textit{forecast model} $F$ with parameters $\theta$, \textit{model adapter} ${MA}$ with parameters $\phi$, and \textit{data adapter} ${DA}$ with parameters $\psi$. 
$DA$ contains a feature adaptation layer $G$ and a label adaptation layer $H$, along with its inverse function $H^{-1}$. In particular, the implementation of $F$ can be realized by any neural network for stock trend forecasting, such as GRU~\cite{GRU} and ALSTM~\cite{ALSTM}.

We cast IL for stock trend forecasting as a sequence of meta-learning tasks.
For each task $(\mathcal D_{\text{train}}^{k}, \mathcal D_{\text{test}}^{k})$, \model involves four steps as listed below.

\begin{enumerate}[leftmargin=*]
    \item \textbf{Incremental data adaptation.} Given incremental data $\mathcal D_{\text{train}}^{k}$, ${DA}$ transforms each feature vector $\mathbf x$ via $G$ and transforms the corresponding label $y$ via $H$, generating an adapted incremental dataset $\mathcal{\widetilde D}_{\text{train}}^{k}$.
    \item\textbf{Model adaptation.\label{step:ma}} ${MA}$ initializes the forecast model by parameter weights $\phi^{k-1}$. Following typical IL, we fine-tune the forecast model on $\widetilde{\mathcal D}_{\text{train}}^{k}$, generating task-specific parameters $\theta^{k}$. Then, the updated forecast model $F(\cdot;{\theta^{k}})$ is deployed online.
    \item\textbf{Online inference.} Given $\mathbf x$ of each sample in $\mathcal D_{\text{test}}^{k}$, $DA$ transforms it via $G$ into $\tilde{\mathbf x}$. Then, $F(\cdot;{\theta^{k}})$ takes $\tilde{\mathbf{x}}$ and produces an intermediate prediction $\check y$ which will be transformed via $H^{-1}$ into the final prediction $\hat y$. We denote the adapted test dataset by $\mathcal{\widetilde D}_{\text{test}}^{k}$ that is obtained by transforming raw features in $\mathcal{D}_{\text{test}}^{k}$.
    \item\textbf{Optimization of meta-learners.\label{step:opt}} We calculate the final forecast error $\mathcal L_{\text{test}}$ once we obtain all ground-truth labels in order to optimize our meta-learners (\textit{i.e.}, ${DA}$ and ${MA}$) in the upper level. The parameters of meta-learners are updated from $\phi^{k-1}$ and $\psi^{k-1}$ to $\phi^{k}$ and $\psi^{k}$, which are used for the next IL task.
\end{enumerate}


The meta-learning process is essentially a bi-level optimization problem, where the fine-tuning in Step (\ref{step:ma}) is the lower-level optimization and Step (\ref{step:opt}) is the upper-level optimization. Formally, since we desire a minimal forecast error on $\mathcal {\widetilde D}_{\text {test}}^{k}$, the bi-level optimization of the $k$-th IL task is defined as

\begin{equation}\label{eq:upper level}
  \phi^{k}, \psi^{k}=\underset{{\phi, \psi}}{\arg \min } \mathcal L_{\text{test}}(\mathcal {\widetilde D}_{\text {test}}^{k}; \theta^{k}),
\end{equation}
\begin{equation}\label{eq:lower level}
  \text {s.t.} \quad {\theta}^{k}={MA}(\mathcal {\widetilde D}_{\text{train}}^{k};\phi^{k-1}), 
\end{equation}
where
\begin{equation}\label{eq:adapted data}
  \mathcal{\widetilde D}_{\text {train }}^{k}={DA}({\mathcal D}_{\text {train }}^{k};\psi^{k-1});\quad
  \mathcal {\widetilde D}_{\text {test}}^{k}={DA}({\mathcal D}_{\text {test}}^{k};\psi^{k-1}).
\end{equation}
Note that we only adapt features of ${\mathcal D}_{\text {test}}^{k}$ because test labels are unknown during online inference time.

Generally, the best incremental learning algorithm should lead to minimum accumulated errors on all test datasets throughout online inference. However, in the context of stock trend forecasting, we only focus on improving the test performance of the current task since past predictions cannot be withdrawn and future datasets are unseen. Therefore, we greedily optimize $\phi$ and $\psi$ task by task in a practical scenario. Furthermore, we approximate Eq.~(\ref{eq:upper level}) by one-step gradient descent for the concern of training efficiency. 

In the following subsections, we elaborate the two adapters and detail our upper-level optimization.

\subsection{Data Adapter}\label{sec:DA}

Data adapter ${DA}$ is a meta-learner to learn what data adaptation is favorable for appropriate updates and promising test performance. ${DA}$ consists of three mapping functions, \textit{i.e.}, feature adaptation $G$, label adaptation $H$, and its inverse mapping $H^{-1}$.


Following our insights, we propose a mapping function $G$ to map $\mathcal P_{\text{train}}^{k}(\mathbf x)$ and $\mathcal P_{\text{test}}^{k}(\mathbf x)$ into an agent feature distribution $\mathcal P_{\text{agent}}^{k}(\mathbf x)$, alleviating covariate shifts. We intuitively introduce a dense layer as a simple implementation of ${G}$, which is defined as follows:
\begin{equation}
  \tilde{\mathbf{x}} = \mathbf W\mathbf x + \mathbf b,
\end{equation}
where $\tilde{\mathbf{x}}$ denotes the adapted feature vector, $\mathbf W\in \mathbb R^{D\times D}$ is the parameter matrix, and $\mathbf b\in \mathbb R^{D}$ is the bias vector. 
Features from ${\mathcal D}_{\text {train }}^{k}$ and ${\mathcal D}_{\text {test}}^{k}$ are thereby transformed onto a new common hyperplane via the same affine transformation. 

The remaining concern is that one simple dense layer is not expressive enough.
Different types of feature vectors may require different transformations, for example, abnormal values need to be scaled or masked, trustworthy features just need identity mapping, and profitable signals need to be emphasized. Moreover, stock price trends tend to bear similar shift patterns when the stocks belong to the same concept (\textit{e.g.}, sector, industry, and business), and vice versa. Accordingly, an appealing solution is to employ different transformation heads and decide which candidate head is more suitable for the input. We thus propose a \textit{multi-head feature adaptation layer} with multiple \textit{feature transformation heads}, which is formally defined as follows:
\begin{equation}\label{eq:feature adaptation}
  \mathbf{\tilde x} = G(\mathbf x) := \mathbf x+\sum_{i=1}^{N} s_i\cdot g_{i}(\mathbf x),
\end{equation}
where we add a residual connection~\cite{ResNet} in case excessive transformation forgets most raw information; $N$ is the number of heads; $g_i$ is the $i$-th feature transformation head; $s_i$ is a normalized confidence score to decide the strength of the $i$-th transformation.
We implement each $g_i$ by a simple dense layer:
\begin{equation}
  g_i(\mathbf x)=\mathbf W_i\mathbf x+\mathbf b_i,  
\end{equation}
where $\mathbf W_i$ and $\mathbf b_i$ are head-specific transformation parameters. 
As for confidence estimation, we use $N$ prototype vectors to prompt which head is more applicable. Specifically, we first calculate the cosine similarity $\hat s_i$ between the feature vector and each prototype, which is formulated as:
\begin{equation}
  \hat s_i=\frac{\mathbf p_i^\top \mathbf x}{\|\mathbf p_i\| \| \mathbf x \|}, 
\end{equation}
where $\mathbf p_i\in \mathbb R^{D}$ is the $i$-th prototype vector. Then, we derive the normalized score $s_i$ by
\begin{equation}
  s_i=\frac{\exp \hat s_i/\tau}{\sum_{j=1}^{N} \exp \hat s_j/\tau},
\end{equation}
where $\tau$ is a positive hyperparameter to control softmax temperature. With the prototypes deemed as learnable embeddings of hidden concepts in the stock market, the multi-head feature adaptation layer can provide concept-oriented transformations for each input feature vector $\mathbf x$.

To cope with conditional distribution shifts, we adapt labels of $\mathcal {D}_{\text {train}}^{k}$. We define a \textit{label transformation head} as follows:
\begin{equation}
  h_i(y) = {\gamma}_i y + \beta_i,
\end{equation}
where ${\gamma}_i\in \mathbb R$ and $\beta_i\in \mathbb R$ are learnable meta-parameters. 
A \textit{multi-head label adaptation layer}, termed as $H$, is formulated as
\begin{equation}\label{eq:label adaptation}
  \tilde{y} = H(y) := \sum_{i=1}^{N} s_i\cdot h_{i}(y),
\end{equation}
where $\tilde y$ denotes the adapted label; $s_i$ has been calculated in the feature adaptation layer and is determined by $\mathbf x$.

During online inference, we need to inversely map the intermediate output $\check{y}$ from $\mathcal P_{\text{agent}}^{k}(y| \mathbf x)$ to $\mathcal P_{\text{test}}^{k}(y| \mathbf x)$, where $\check y = F(\mathbf x)$. We define the inverse mapping function $H^{-1}$ by
\begin{equation}
  \hat y = H^{-1}(\mathbf x, \check y) := \sum_{i=1}^{N} s_i\cdot h^{-1}_{i}(\check y),
\end{equation}
where 
\begin{equation}
  h_i^{-1}(\check y) = ({\check y - \beta_i})/{\gamma_i}.
\end{equation}
One can seek further improvements on the implementation of each label adaptation head $h_i$ by various normalizing flows~\cite{Normalizing_Flow_Survey}, which comprise a sequence of invertible mappings. Empirically, we show that the simple linear heads have already been effective in Sec.~\ref{sec:ablation}.

To summarize, meta-parameters $\psi$ of the data adapter ${DA}$ include the parameters of $G$ and $H$, \textit{i.e.}, $\psi = \{\mathbf W_i, \mathbf b_i, \mathbf p_i, \gamma_i, \beta_i\}_{i=1}^{N}$. The data adapter transforms features of $\mathcal {D}_{\text {train}}^{k}$ and $\mathcal {D}_{\text {test}}^{k}$ by Eq.~(\ref{eq:feature adaptation}), and transforms labels of $\mathcal {D}_{\text {train}}^{k}$ by Eq.~(\ref{eq:label adaptation}). The adapted datasets $\mathcal {\widetilde D}_{\text {train}}^{k}$ and $\mathcal{\widetilde D}_{\text {test}}^{k}$ in Eq.~(\ref{eq:adapted data}) can be formalized as
\begin{subequations}\label{eq:final adapted data}
  \begin{align}
  \mathcal{\widetilde D}_{\text {train}}^{k}&=\{(G(\mathbf x), H(\mathbf x, y))\mid (\mathbf x, y)\in \mathcal{D}_{\text {train }}^{k}\},\label{eq:final adapted training data}\\ 
  \mathcal{\widetilde D}_{\text {test }}^{k}&=\{(G(\mathbf x), y)\mid (\mathbf x, y)\in \mathcal{D}_{\text {test}}^{k}\}\label{eq:final adapted test data},
  \end{align}
\end{subequations}
where the ground-truth labels in $\mathcal{\widetilde D}_{\text {test}}^{k}$ are known at the corresponding trading dates in the future.

\subsection{Model Adapter}

Model adapter ${MA}$ is another meta-learner to provide a good initialization of model parameters for each IL task and then adapt the initial parameters to fit the incremental data. For the $k$-th IL task, ${MA}$ first assigns the forecast model $F$ with initial weights $\phi^{k-1}$, and then fine-tunes $F({\cdot;\phi^{k-1}})$ on $\mathcal{\widetilde D}_{\text{train}}^{k}$. The optimization objective is to reduce the training loss $\mathcal{L}_{\text{train}}$ which is defined as:

\begin{equation}\label{eq:training loss}
  \mathcal{L}_{\text{train}}=\frac{1}{|\mathcal{\widetilde D}_{\text{train}}^{k}|}\sum_{(\tilde{\mathbf x},\tilde y)\in \mathcal{\widetilde D}_{\text{train}}^{k}} \left(F(\tilde{\mathbf x};\phi^{k-1})-\tilde y\right)^2.
\end{equation}
We adapt the initial parameters $\phi^{k-1}$ and derive task-specific parameters $\theta^{k}$ as follows:
\begin{equation}\label{eq:task-specific param}
  \begin{aligned}  
  {\theta}^{k}&={MA}(\mathcal {\widetilde D}_{\text{train}}^{k};\phi^{k-1})\\
                &=\phi^{k-1} -\eta_\theta \nabla_\phi \mathcal{L}_{\text{train}}(\mathcal {\widetilde D}_{\text {train}}^{k}; \phi^{k-1}),
  \end{aligned}
\end{equation}
where $\eta_\theta$ is the learning rate of the forecast model and we only perform one-step gradient updates for fast adaptation. Next, we deploy $F(\cdot;{\theta^{k}})$ online to make predictions on adapted test data $\mathcal {\widetilde D}_{\text{test}}^{k}$. 

Note that \model is a generic framework for any implementation of the model adapter. 
The one-step gradient update can also be replaced by other model adaptation approaches, such as context-dependent gating~\cite{context-dependent-gating} and transfer network~\cite{SML}.

\subsection{Optimization of Meta-learners}\label{sec:optimization}
At each test time, we can calculate the test error once the ground-truth labels of the test data are known. Formally, we obtain mean square error $\mathcal L_{\text{MSE}}$ to optimize our meta-learners before we start the next task, which is defined as follows:
\begin{equation}
  \mathcal L_{\text{MSE}}=\frac{1}{|\mathcal{\widetilde D}_{\text {test}}^{k}|}\sum_{(\tilde{\mathbf x}, y) \in \mathcal {\widetilde D}_{\text {test}}^{k}}\left( H^{-1}\circ F(\tilde{\mathbf x} ; \theta^{k}, \psi^{k-1}) - y\right)^2.
\end{equation}
Additionally, we add a regularization term to avoid the adapted labels of $\mathcal{\widetilde D}_{\text {train}}^{k}$ being abnormal and hard to learn. 
We formulate the regularization loss $\mathcal L_{\text{reg}}$ by
\begin{equation}
  \mathcal L_{\text{reg}}=\frac{1}{|\mathcal{D}_{\text {train}}^{k}|}\sum_{(\mathbf{x}, y) \in \mathcal {D}_{\text {train}}^{k}}\left( H(\mathbf{\tilde x}, y; \psi^{k-1}) - y\right)^2.
\end{equation}
The final test loss $\mathcal L_{\text{test}}$ is derived by
\begin{equation}\label{eq:test loss}
  \mathcal L_{\text{test}}=\mathcal L_{\text{MSE}}+\alpha\mathcal L_{\text{reg}},
\end{equation}
where $\alpha$ is a hyperparameter to control the regularization strength. 
Actually, $\alpha\mathcal L_{\text{reg}}$ approximates the second-order gradients of the label adaptation layer (see derivations in Appendix ~\ref{apx:approximation}).

The optimization of our model adapter ${MA}$ is formulated as:
\begin{equation}
  {\phi}^{k} = {\phi}^{k-1}-\eta_\phi \nabla_\phi \mathcal{L}_{\text{test}}(\mathcal {\widetilde D}_{\text {test}}^{k}; \theta^{k}),
\end{equation}
where $\eta_\phi$ is the learning rate of ${MA}$. The difference between traditional MAML~\cite{MAML} and our method lies in that we only use one query set at each time to update the meta-learner. We are devoted to improving performance on the current test data because the stock market evolves and a large proportion of past data may not contain patterns that will reappear in the future. It is also practical in the IL setting where we only save samples of one task in memory.

Similarly, we optimize ${DA}$ according to actual test performance, using the following equation:
\begin{equation}
  {\psi}^{k} = {\psi}^{k-1}-\eta_\psi \nabla_\psi \mathcal{L}_{\text{test}}(\mathcal {\widetilde D}_{\text {test}}^{k}; \theta^{k}),
\end{equation}
where $\eta_\psi$ is the learning rate of ${DA}$. $\mathcal {\widetilde D}_{\text {test}}^{k}$ and $\theta^{k}$ are derived by Eq.~(\ref{eq:final adapted training data}) and Eq.~(\ref{eq:task-specific param}), respectively.

\section{Training Procedure} \label{sec:procedure}
In this section, we propose a two-phase training procedure of \model including an offline training phase and an online training phase. Pseudo-codes are shown in Alg.~\ref{alg:overall} and Alg.~\ref{alg:task}.

\begin{algorithm}[b]
  \KwIn{Meta-train set $\mathcal T_{\text{train}}=\{(\mathcal D_{\text {train }}^{k}, \mathcal D_{\text {test}}^{k})\}_{k=1}^{K_0}$, meta-valid set $\mathcal T_{\text{valid}}=\{(\mathcal D_{\text {train }}^{k}, \mathcal D_{\text {test}}^{k})\}_{k=K_0+1}^{K}$ and meta-test set $\mathcal T_{\text{test}}=\{(\mathcal D_{\text {train }}^{k}, \mathcal D_{\text {test}}^{k})\}_{k=K+1}^{K^\prime}$
  }
  \KwOut{Trained parameters $\phi$, $\psi$ and online predictions $\{\mathbf {\hat Y}^{(t)}\}_{t=T+1}^{T^\prime}$}
  \BlankLine
  \caption{Overall training procedure}\label{alg:overall}

  Initialize $\phi$ and $\psi$\;  
  
  \tcc{Offline training phase}
  \Repeat{\textnormal{the metric decreases for $\zeta$ epochs}}{
    $\mathcal T_{\text{train}}^\prime \leftarrow \mathtt{shuffle}(\mathcal T_{\text{train}})$\;
    $\phi$, $\psi$, $\{\hat{\mathbf Y}^{(t)}\}_{\scriptscriptstyle t=r+1}^{\scriptscriptstyle (K_0+1)r}$$\leftarrow \mathtt{\modelns}(\phi, \psi, \mathcal T_{\text{train}}^\prime)$\;
    ${\phi^\prime}$, ${\psi^\prime} \leftarrow$ make a temporary copy of $\phi$ and $\psi$\label{alg:overall:copy}\;
    ${\phi^\prime}$, ${\psi^\prime}$, $\{\hat{\mathbf Y}^{(t)}\}_{\scriptscriptstyle t=(K_0+1)r+1}^{\scriptscriptstyle (K+1)r}$$\leftarrow \mathtt{\modelns}({\phi^\prime}, {\psi^\prime}, \mathcal T_{\text{valid}})$\label{alg:overall:eval}\;
    Calculate metric on $\{\hat {\mathbf Y}^{(t)}, {\mathbf Y}^{(t)}\}_{\scriptscriptstyle t=(K_0+1)r+1}^{\scriptscriptstyle (K+1)r}$\;
  } 
  \BlankLine
  \tcc{Online training phase from $T+1$ to $T^\prime$}
  $\phi$, $\psi$, $\{\hat{\mathbf Y}^{(t)}\}_{\scriptscriptstyle t=H+(K+1)r}^{\scriptscriptstyle T'}$$\leftarrow \mathtt{\modelns}(\phi, \psi, \mathcal T_{\text{valid}}\cup \mathcal T_{\text{test}})$\label{alg:overall:online}\; 
  Evaluate \model by metric on $\{\hat{\mathbf Y}^{(t)}, {\mathbf Y}^{(t)}\}_{t=T+1}^{T^\prime}$\;

  \Return $\phi$, $\psi$, $\{\hat{\mathbf Y}^{(t)}\}_{t=T+1}^{T^\prime}$
\end{algorithm}

\begin{algorithm}[t]
  \KwIn{$\phi$, $\psi$ and task set $\mathcal T=\{(\mathcal D_{\text {train }}^{k}, \mathcal D_{\text {test}}^{k})\}_{k=m}^{n}$}
  \KwOut{$\phi^{n}$, $\psi^{n}$ and predictions $\{\hat{\mathbf Y}^{(t)}\}_{t=mr+1}^{(n+1)r}$}
  \BlankLine
  \caption{\modelns}\label{alg:task}

  $\phi^{(m-1)}, \psi^{(m-1)} \leftarrow \phi$, $\psi$\;  
  \For{$k \leftarrow m$ \KwTo $n$}{
    Assign forecast model $F$ with inital weights $\phi^{k-1}$\;

    Predict on $\widetilde{\mathcal D}_{\text {train }}^{k}$ defined by Eq.~(\ref{eq:final adapted training data})\;

    Compute training loss by Eq.~(\ref{eq:training loss})\;
    
    Adapt the forecast model parameters by:
    $$
    {\theta}^{k}=\phi^{k-1}-\eta_\theta \nabla_\phi \mathcal{L}_{\text{train}}(\mathcal {\widetilde D}_{\text {train}}^{k}; \phi^{k-1});\quad
    $$

    $\{\hat{\mathbf Y}^{(t)}\}_{t=kr+1}^{(k+1)r}$$\leftarrow$ Predict on $\widetilde{\mathcal D}_{\text {test}}^{k}$ defined by Eq.~(\ref{eq:final adapted test data})

    Compute test loss by Eq.~(\ref{eq:test loss})\;
    
    Update data adapter ${DA}$ and model adapter ${MA}$:\label{alg:task:upper}
    $$
    {\psi}^{k} = {\psi}^{k-1}-\eta_\psi \nabla_\psi \mathcal{L}_{\text{test}}(\mathcal {\widetilde D}_{\text {test}}^{k}; \theta^{k}),\quad\quad\quad
    $$
    $$
    {\phi}^{k} = {\phi}^{k-1}-\eta_\phi \nabla_\phi \mathcal{L}_{\text{test}}(\mathcal {\widetilde D}_{\text {test}}^{k}; \theta^{k}).\quad\quad\quad
    $$
  }
  \Return $\phi^{n}$, $\psi^{n}$, $\{\hat{\mathbf Y}^{(t)}\}_{t=mr+1}^{(n+1)r}$
\end{algorithm}

Given all historical data $\{(\mathbf X^{(t)}, \mathbf Y^{(t)})\}_{t=1}^{T}$ before deployment, we organize these offline data into $K$ incremental learning tasks to imitate online incremental learning. The $k$-th IL task consists of the $k$-th incremental data $\mathcal D_{\text{train}}^{k}=\{(\mathbf{X}^{(t)}, \mathbf Y^{(t)})\}_{t=(k-1)r+1}^{kr}$ and the $k$-th test data $\mathcal D_{\text{test}}^{k}=\{(\mathbf{X}^{(t)}, \mathbf Y^{(t)})\}_{t=kr+1}^{(k+1)r}$. We take the first $K_0$ tasks as meta-train set $\mathcal T_{\text{train}}$ and others as meta-valid set $\mathcal T_{\text{valid}}$. It is noteworthy that we can shuffle $\mathcal T_{\text{train}}$ to simulate arbitrary distribution shifts in pursuit of robustness against extreme non-stationarity.

\vspace{0.3em}
\noindent\textbf{Offline training.} We pretrain the two meta-learners on $\mathcal T_{\text{train}}$ task by task for epochs. At the end of each epoch, we continue incremental learning on $\mathcal T_{\text{valid}}$ for evaluation (Alg.~\ref{alg:overall}, L\ref{alg:overall:eval}). The updates of the meta-learners on the meta-valid set are conducted on a temporary copy of the meta-learners (Alg.~\ref{alg:overall}, L\ref{alg:overall:copy}), in order to avoid information leakage. We perform early stopping when the evaluation metric on $\mathcal T_{\text{valid}}$ decreases for $\zeta$ consecutive epochs, in order to avoid overfitting. The meta-learners get ready for online service after they have been incrementally updated on the whole validation set (Alg.~\ref{alg:overall}, L\ref{alg:overall:online}).

\vspace{0.3em}
\noindent\textbf{Online training.} 
We deploy \model online at date $T$ and incrementally update the meta-learners at the end of each online task (Alg.~\ref{alg:task}, L\ref{alg:task:upper}). As a special case, the meta-learners could keep the original adaptation ability of traditional MAML when the learning rates $\eta_\psi$ and $\eta_\phi$ are always set to zero during the online training phase. However, as the stock market is dynamically evolving, it is critical to continually consolidate the meta-learners with new knowledge, which is also empirically confirmed by our experimental results in Appendix~\ref{sec:hyperparam}.

In case the meta-learners encounter the catastrophic forgetting problem, we can also restart offline training after a much larger interval. For example, we incrementally update the meta-learners every week and fully retrain them on an enlarged meta-train set after one-year incremental learning. Integrating the advantages of IL and RR, \model is expected to achieve better performance. In this work, we focus on improving IL performance against distribution shifts and leave the combination of \model and RR as future work. According to our experiments on two real-world stock datasets, \model alone can outperform RR methods for a long period (\textit{e.g.}, over 2.5 years) and hence it is unnecessary to frequently perform full retraining.

\vspace{0.3em}
\noindent\textbf{Complexity analysis.} When optimizing the meta-learners (Alg.~\ref{alg:task}, L\ref{alg:task:upper}), we adopt the first-order approximation version of MAML ~\cite{MAML} to avoid the expensive computation of Hessian matrices. Therefore, both the time and memory costs per IL task are linearly proportional to the size of incremental data. Formally, let $S$ denote the number of stocks in the market and $E$ denote the number of training epochs for RR methods till convergence. For the $k$-th online task, both the incremental data and the test data have $rS$ samples, while RR takes $(T+kr)S$ historical samples to train the model from scratch. Hence, the time complexity of DoubleAdapt and RR is $O(rS)$ and $O(E(T+kr)S)$, respectively. The scalability of \model allows frequent updates on the forecast model, \textit{e.g.}, on a weekly basis.

\section{Experiments}

In this section, we study our \model framework with experiments, aiming to answer the following research questions:
\begin{itemize}[leftmargin=*]
\item \textbf{RQ1}: How does our proposed \model approach perform compared with the state-of-the-art methods?
\item \textbf{RQ2}: How is the effect of different components in \modelns?
\item \textbf{RQ3}: What is the empirical time cost of \modelns?
\end{itemize}
An additional hyperparameter study is provided in Appendix~\ref{sec:hyperparam}.

\subsection{Experimental Settings}
\subsubsection{Datasets}
We evaluate our \model framework on two popular real-world stock sets: CSI 300~\cite{REST, Yoo2021, CMLF, HIST} and CSI 500~\cite{REST, DDGDA} in the China A-share market. CSI 300 consists of the 300 largest stocks, reflecting the overall performance of the market. CSI 500 comprises the largest remaining 500 stocks after excluding the CSI 300 constituents, reflecting the small-mid cap stocks.

We use the stock features of Alpha360 in the open-source quantitative investment platform Qlib~\cite{Qlib}. Alpha360 contains 6 indicators on each day, which are \textit{opening price, closing price, highest price, lowest price, volume weighted average price} (VWAP) and \textit{trading volume}. For each stock at date $t$, Alpha360 looks back 60 days to construct a 360-dimensional vector as the raw feature of this stock. We use the stock price trend defined in Definition~\ref{def:trend} as the label for each stock. 
Following Qlib, we split stock data into training set (from 01/01/2008 to 12/31/2014), validation set (from 01/01/2015 to 12/31/2016), and test set (from 01/01/2017 to 07/31/2020). The features are normalized by moments of the whole training set, and the labels grouped by date are normalized by moments of data at the same date~\cite{REST, HIST}.

\subsubsection{Evaluation Metrics}
We use four widely-used evaluation metrics: IC~\cite{TRA}, ICIR~\cite{AdaRNN}, Rank IC~\cite{Li2019b}, and Rank ICIR~\cite{HIST}. At each date $t$, $IC^{(t)}$ could be measured by
\begin{equation}
    IC^{(t)} = \frac{1}{N}\frac{(\hat{\mathbf{Y}}^{(t)} - mean(\hat{\mathbf{Y}}^{(t)}))^{\top} (\mathbf{Y}^{(t)} - mean(\mathbf{Y}^{(t)}))}{std(\hat{\mathbf{Y}}^{(t)})\cdot std(\mathbf{Y}^{(t)})},
\end{equation}
where ${\mathbf{Y}^{(t)}}$ are the raw stock price trends and $\hat{\mathbf{Y}}^{(t)}$ are the model predictions at each date. We report the average IC over all test dates.
ICIR is calculated by dividing the average by the standard deviation of IC. Rank IC and Rank ICIR are calculated by ranks of labels and ranks of predictions.
Besides, we also use two portfolio metrics, including the excess annualized return (Return) and its information ratio (IR). IR is calculated by dividing the excess annualized return by its standard deviation. 
Our backtest settings adhere to Qlib's default strategy. 

We ran each experiment 10 times and report the average results. 
For all six metrics, a higher value reflects better performance.

\subsubsection{Baseline}
We consider two kinds of model-agnostic retraining approaches as the comparison methods:

\vspace{0.3em}
\noindent\textit{(1) Rolling retraining methods}:
\begin{itemize}[leftmargin=*]
\item \textbf{RR}~\cite{DDGDA}: RR, short for rolling retraining, periodically retrains a model on all available data with equal weights. 
\item \textbf{DDG-DA}~\cite{DDGDA}: This method predicts the data distribution of the next time-step sequentially and re-weights all historical samples to generate a training set, of which the distribution is similar to the predicted future distribution. 
\end{itemize}
\textit{(2) Incremental learning methods}:
\begin{itemize}[leftmargin=*]
\item \textbf{IL}: This method is a na\"ive incremental learning baseline to fine-tune the model only with the recent incremental data by gradient descent~\cite{OGD}.
\item \textbf{MetaCoG}~\cite{CML}: This method introduces a per-parameter mask to select task-specific parameters according to the context. 
MetaCoG updates the masks rather than the model parameters to avoid catastrophic forgetting.
\item \textbf{C-MAML}~\cite{OSAKA}: This method follows MAML~\cite{MAML} to pretrain slow weights that can produce fast weights to accommodate new tasks. At the online time, C-MAML keeps fine-tuning the fast weights until a distribution shift is detected, and then the slow weights are updated and used to initialize new fast weights.
\item \textbf{\modelns}: Our proposed method. \model learns to initialize parameters and, notably, learns to adapt features and labels, alleviating the distribution shift issue.
\end{itemize}
Note that there are few studies on incremental learning for stock trend forecasting. We borrow MetaCoG and C-MAML from the continual learning problem as IL-based baselines.

\subsubsection{Implementation Details}\label{sec:details}
The time interval of two consecutive tasks is 20 trading days~\cite{HIST}. The batch size of RR-based methods approximates the number of samples in incremental data, \textit{i.e.}, 5000 for CSI 300 and 8000 for CSI 500. We apply Adam optimizer with an initial learning rate of 0.001 for the forecast model of all baselines, 0.001 for our model adapter, and 0.01 for our data adapter. We perform early stopping when IC decreases for 8 consecutive epochs. The regularization strength $\alpha$ of \model is 0.5. The head number $N$ is 8 and the temperature $\tau$ is 10. Other hyperparameters of the forecast model (\textit{e.g.}, dimension of hidden states) keep the same for a fair comparison. We use the \textbf{first-order approximation} version of MAML~\cite{MAML} for all meta-learning methods. 

\begin{table*}[t]
	\centering
	\caption{Overall performance comparison on CSI 300 and CSI 500, where the bold values are the best results and the underlined values are the most competitive results (RQ1).
 }\label{tab:main_result}
  \resizebox{\textwidth}{!}{
\begin{tabular}{cc|cccccc|cccccc}
\toprule
\multirow{2}{*}{\textbf{Model}} & \multirow{2}{*}{\textbf{Method}} & \multicolumn{6}{c|}{\textbf{CSI 300}} & \multicolumn{6}{c}{\textbf{CSI 500}} \\
 &  & \textbf{\quad IC \quad} & \textbf{\quad ICIR\quad } & \textbf{RankIC} & \textbf{RankICIR} & \textbf{Return} & \textbf{\quad \ IR\ \quad } & \textbf{\quad IC\quad } & \textbf{\quad ICIR\quad } & \textbf{RankIC} & \textbf{RankICIR} & \textbf{Return} & \textbf{\quad \ IR\ \quad } \\\midrule
 \multirow{6}{*}{\makecell{Trans-\\former}} & RR & 0.0449 & 0.3410 & \underline{0.0462} & \textbf{0.3670} & 0.0881 & 1.0428 & 0.0452 & 0.4276 & \underline{0.0469} & 0.4732 & 0.0639 & 0.9879 \\
 & DDG-DA & 0.0420 & 0.3121 & 0.0441 & 0.3420 & 0.0823 & 1.0018 & 0.0450 & 0.4223 & 0.0465 & 0.4634 & 0.0681 & 1.0353 \\
 & IL & 0.0431 & 0.3108 & 0.0411 & 0.2944 & 0.0854 & 0.9215 & 0.0428 & 0.3943 & 0.0453 & 0.4475 & 0.1014 & \underline{1.5108} \\
 & MetaCoG & 0.0463 & 0.3493 & 0.0434 & 0.3133 & 0.0952 & 0.9921 & 0.0449 & \underline{0.4643} & 0.0469 & 0.4629 & \underline{0.1053} & 0.8945 \\
  & C-MAML & \underline{0.0479} & \underline{0.3560} & 0.0448 & 0.3405 & \underline{0.0986} & \underline{1.0537} & \underline{0.0477} & 0.4620 & 0.0468 & \underline{0.4861} & 0.0930 & 1.4923 \\
 & DoubleAdapt & \textbf{0.0516} & \textbf{0.3889} & \textbf{0.0475} & \underline{0.3585} & \textbf{0.1041} & \textbf{1.1035} & \textbf{0.0492} & \textbf{0.4653} & \textbf{0.0490} & \textbf{0.4970} & \textbf{0.1330} & \textbf{1.9761} \\\midrule
\multirow{6}{*}{LSTM} & RR & 0.0592 & \underline{0.4809} & 0.0536 & \underline{0.4526} & 0.0805 & 0.9578 & \underline{0.0642} & \underline{0.6187} & 0.0543 & 0.5742 & 0.0980 & 1.5220 \\
 & DDG-DA & 0.0572 & 0.4622 & 0.0528 & 0.4415 & 0.0887 & 1.0583 & 0.0636 & 0.6181 & 0.0540 & 0.5783 & 0.1061 & 1.6673 \\
 & IL & \underline{0.0594} & 0.4664 & \underline{0.0546} & 0.4362 & \underline{0.1089} & \underline{1.2553} & 0.0576 & 0.5550 & \underline{0.0553} & 0.5660 & 0.1249 & 1.8461 \\
 & MetaCoG & 0.0515 & 0.4131 & 0.0505 & 0.4197 & 0.1013 & 1.1133 & 0.0573 & 0.5673 & 0.0549 & \underline{0.5908} & \underline{0.1384} & \underline{2.0546} \\
 & C-MAML & 0.0568 & 0.4601 & 0.0517 & 0.4381 & 0.0963 & 1.1145 & 0.0582 & 0.5863 & 0.0550 & 0.5898 & 0.1315 & 1.9770 \\
 & DoubleAdapt & \textbf{0.0632} & \textbf{0.5126} & \textbf{0.0567} & \textbf{0.4669} & \textbf{0.1117} & \textbf{1.3029} & \textbf{0.0648} & \textbf{0.6331} & \textbf{0.0594} & \textbf{0.6087} & \textbf{0.1496} & \textbf{2.2220} \\\midrule
\multirow{6}{*}{ALSTM} & RR & 0.0630 & \underline{0.5084} & \underline{0.0589} & \textbf{0.4892} & 0.0947 & 1.1785 & \underline{0.0649} & 0.6331 & 0.0575 & 0.6030 & 0.1211 & 1.8726 \\
 & DDG-DA & 0.0609 & 0.4915 & 0.0581 & 0.4823 & 0.0966 & 1.2227 & 0.0645 & 0.6298 & 0.0573 & 0.6029 & 0.1042 & 1.6091 \\
 & IL & 0.0626 & 0.4762 & 0.0585 & 0.4489 & \underline{0.1171} & \underline{1.3349} & 0.0596 & 0.5705 & 0.0579 & 0.5712 & 0.1501 & 2.1468 \\
 & MetaCoG & 0.0581 & 0.4676 & 0.0570 & 0.4695 & 0.1140 & 1.3228 & 0.0576 & 0.5874 & 0.0571 & 0.6086 & 0.1403 & 2.0857 \\
 & C-MAML & \underline{0.0636} & 0.5064 & 0.0588 & 0.4765 & 0.1085 & 1.2432 & 0.0647 & \textbf{0.6490} & \underline{0.0598} & \textbf{0.6330} & \underline{0.1644} & \underline{2.4636} \\
 & DoubleAdapt & \textbf{0.0679} & \textbf{0.5480} & \textbf{0.0594} & \underline{0.4882} & \textbf{0.1225} & \textbf{1.4717} & \textbf{0.0653} & \underline{0.6404} & \textbf{0.0607} & \underline{0.6170} & \textbf{0.1738} & \textbf{2.5192} \\\midrule
\multirow{6}{*}{GRU} & RR & 0.0629 & \underline{0.5105} & 0.0581 & 0.4856 & 0.0933 & 1.1428 & \underline{0.0669} & \underline{0.6588} & 0.0586 & 0.6232 & 0.1200 & 1.8629 \\
 & DDG-DA & 0.0623 & 0.5045 & 0.0589 & \underline{0.4898} & 0.0967 & 1.1606 & 0.0666 & 0.6575 & 0.0582 & 0.6234 & 0.1264 & 1.9963 \\
 & IL & 0.0633 & 0.4818 & \underline{0.0596} & 0.4609 & \underline{0.1166} & 1.3196 & 0.0637 & 0.6093 & \underline{0.0617} & 0.6291 & 0.1626 & 2.3352 \\
 & MetaCoG & 0.0560 & 0.4443 & 0.0545 & 0.4503 & 0.0992 & 1.1014 & 0.0603 & 0.5741 & 0.0585 & 0.5720 & 0.1587 & 2.2635 \\
 & C-MAML & \underline{0.0638} & 0.5085 & 0.0595 & 0.4865 & 0.1121 & \underline{1.3210} & 0.0646 & 0.6498 & 0.0600 & \textbf{0.6494} & \underline{0.1693} & \textbf{2.5064} \\
 & DoubleAdapt & \textbf{0.0687} & \textbf{0.5497} & \textbf{0.0621} & \textbf{0.5110} & \textbf{0.1296} & \textbf{1.5123} & \textbf{0.0686} & \textbf{0.6652} & \textbf{0.0632} & \underline{0.6445} & \textbf{0.1748} & \underline{2.4578} \\
 \bottomrule
  \end{tabular}
  }
\end{table*}

\subsection{Performance Comparison (RQ1)}

We instantiate the forecast model by four deep neural networks, including Transformer~\cite{Transformer}, LSTM~\cite{LSTM}, ALSTM~\cite{ALSTM}, and GRU~\cite{GRU}. Table~\ref{tab:main_result} compares the overall performance of all baselines. Though RR methods are strong baselines and often beat simple IL methods, our proposed \model framework achieves the best results in almost all the cases, demonstrating that \model can make more precise predictions in stock trend forecasting. Exceptionally, C-MAML sometimes achieves higher ICIR or Rank ICIR than \model on CSI 500. During the online training phase, C-MAML modulates the learning rate of its meta-learner according to the test error and thus achieves more stable performance on daily IC. Note that this update modulation is orthogonal to our work and can also be integrated into \modelns. We also observe that MetaCoG is nearly the worst method that even performs worse than na\"ive IL. MetaCoG focuses on catastrophic forgetting issues and selectively masks the model parameters, instead of consolidating the parameters with new knowledge acquired online. This observation confirms the significance of online updates for stock trend forecasting. 
As for the implementation of the forecast model, \model can consistently achieve better performance with a stronger backbone.

\begin{table*}[t]
	\centering
	\caption{Ablation study on CSI 300 (RQ2). The forecast model is GRU. The head number $N$ is 8 unless otherwise stated. $IL+{MA}+{DA}$ is our proposed method. The column \textit{Overall Performance} corresponds to the results on CSI 300 in Table~\ref{tab:main_result}.
 }\label{tab:ablation}
  \resizebox{\linewidth}{!}{
  \begin{tabular}{l|cccc|cccc|cccc}
\toprule
\multicolumn{1}{c|}{\multirow{2}{*}{\textbf{Method}}} & \multicolumn{4}{c|}{\textbf{Overall Performance}}                                        & \multicolumn{4}{c|}{\textbf{Gradual Shifts}}                                   & \multicolumn{4}{c}{\textbf{Abrupt Shifts}}                                         \\
\multicolumn{1}{c|}{}                                 & \textbf{IC}     & \textbf{ICIR}   & \textbf{RankIC} & \textbf{RankICIR} & \textbf{IC}     & \textbf{ICIR}   & \textbf{RankIC} & \textbf{RankICIR} & \textbf{IC}     & \textbf{ICIR}   & \textbf{RankIC} & \textbf{RankICIR} \\\midrule
IL & 0.0633 & 0.4818 & 0.0596 & 0.4609 & 0.0643 & 0.4936 & 0.0652 & 0.5161 & 0.0690 & 0.5134 & 0.0619 & 0.4581 \\
$+{DA}$ & 0.0659 & 0.5279 & 0.0615 & 0.4993 & 0.0708 & 0.5938 & 0.0692 & 0.5897 & 0.0690 & 0.5271 & 0.0620 & 0.4677 \\
$+{MA}$ & 0.0658 & 0.5160 & 0.0610 & 0.4910 & 0.0703 & 0.5703 & 0.0680 & 0.5686 & 0.0681 & 0.5085 & 0.0618 & 0.4594 \\
$+{MA}$+$G$ & 0.0678 & 0.5360 & 0.0619 & 0.4978 & 0.0740 & 0.6155 & \underline{0.0709} & \underline{0.6060} & 0.0694 & 0.5224 & 0.0626 & 0.4672 \\
$+{MA}$+$H$+$H^{-1}$ & 0.0660 & 0.5207 & 0.0614 & 0.4995 & 0.0714 & 0.5846 & 0.0701 & 0.5958 & 0.0680 & 0.5093 & 0.0616 & 0.4615  \\
$+{MA}$+${DA}$ ($N$=1) & \underline{0.0684} & \underline{0.5462} & \textbf{0.0622} & \underline{0.5074} & \underline{0.0744} & \underline{0.6227} & 0.0703 & 0.6029 & \textbf{0.0710} & \textbf{0.5430} & \textbf{0.0634} & \textbf{0.4822} \\
\textbf{+\textit{MA}+\textit{DA} ($\boldsymbol{N}$=8)} & \textbf{0.0687} & \textbf{0.5497} & \underline{0.0621} & \textbf{0.5110} & \textbf{0.0755} & \textbf{0.6390} & \textbf{0.0713} & \textbf{0.6243} & \underline{0.0699} & \underline{0.5323} & \underline{0.0620} & \underline{0.4730}\\
\bottomrule
\end{tabular}
}
\end{table*}

\subsection{Ablation Study (RQ2)}\label{sec:ablation}

In this section, we investigate the effects of the key components in \modelns. Note that IL is also a special case of \model when ${DA}$ always provides identity mappings and $\phi^{k-1}$ is directly updated into $\theta^{k}$ instead of performing gradient descent. We introduce some variants by equipping IL with one or several components. Besides, as the performance varies in different degrees of distribution shifts, we also evaluate the online predictions under different kinds of shifts. To this end, we pretrain the model $F$ on the training set and adopt na\"ive IL throughout the tasks. In the $k$-th task of the test set, we first use $F$ to do inference on $\mathcal D^{k}_{\text{test}}$ without training on $\mathcal D^{k}_{\text{train}}$, resulting in a mean square error $\mathcal L^{k}_{1}$. Then we update $F$ on $\mathcal D^{k}_{\text{train}}$ and infer the same test samples, resulting in a new mean square error $\mathcal L^{k}_{2}$. The distribution shift in this task can be measured by $\Delta \mathcal L^{k} = \mathcal L^{k}_2 - \mathcal L^{k}_1$ which can reflect whether $F$ benefits from the incremental update. When $\mathcal P^{k}_{\text{test}}$ is similar to $\mathcal P^{k}_{\text{train}}$, $\mathcal L^{k}_2$ should be smaller than $\mathcal L^{k}_1$, and vice versa. With all online tasks sorted by $\Delta \mathcal L^{k}$ in an ascending order, we take the first 25\% tasks as cases of gradual shifts and the last 25\% tasks as cases of abrupt shifts.

Table~\ref{tab:ablation} shows the average results over the two kinds of distribution shifts on CSI 300, \textit{i.e.}, gradual shifts and abrupt shifts. Applying both ${MA}$ and ${DA}$ into IL, \model achieves the best results against different kinds of distribution shifts.
In the cases of gradual shifts, the multi-head version with better expressiveness significantly outperforms the single-head one. Nevertheless, the multi-head version performs worse in the cases of abrupt shifts because its high complexity is a double-edged sword and may incur overfitting issues. As gradual shift is the major issue in stock data~\cite{DDGDA}, the multi-head version still achieves the best overall performance. 
Also, it is noteworthy that our proposed data adaptation effectively facilitates model adaptation, and data adaptation alone also beats one-sided model adaptation. On the other hand, the improvement of \model over $IL$+$DA$, especially under abrupt shifts, indicates that the initial parameters of each task are also critical to generalization ability. 

Moreover, either the data adaptation or the model adaptation outperforms the most competitive methods (\textit{i.e.,} DDG-DA and C-MAML) in Table~\ref{tab:main_result}. Specifically, IL+DA outperforms DDG-DA by 5.6\% improvement on IC, and IL+MA outperforms C-MAML by 3.3\% improvement on IC. In terms of model adaptation, C-MAML proposes additional modules to deal with catastrophic forgetting in general continual learning problems. However, future stock trends are mainly affected by recent stock trends, and it is more beneficial to learn new patterns from recent data rather than memorize long-term historical patterns. Thus, our simple variant IL+MA achieves better performance than C-MAML.


\subsection{Time Cost Study (RQ3)}
\begin{table}[t]
	\centering
	\caption{Empirical time cost (in second) comparison.}\label{tab:time cost}
  \resizebox{0.95\linewidth}{!}{
\begin{tabular}{cccccc}
\toprule
\multirow{2}{*}{\textbf{Model}} & \multirow{2}{*}{\textbf{Method}} & \multicolumn{2}{c}{\textbf{CSI 300}} & \multicolumn{2}{c}{\textbf{CSI 500}} \\
                                &                                   & \textbf{Offline}  & \textbf{Online} & \textbf{Offline}  & \textbf{Online} \\
                                \midrule
\multirow{6}{*}{GRU}            & RR                                & -                 & 6064            & -                 & 10793           \\
                                & DDG-DA                            & 1862              & 6719            & 2360              & 10713           \\
                                & IL                                & \textbf{256}      & \textbf{58}     & \textbf{394}      & \textbf{75}     \\
                                & MetaCoG                           & 457               & 59              & 784               & 76              \\
                                & C-MAML                            & 314               & 62              & 533               & 77             \\
                                & \textbf{DoubleAdapt}                              & 356               & 61              & 677               & 79            \\\bottomrule
\end{tabular}
}
\end{table}

Table~\ref{tab:time cost} compares the total time costs in offline training and online training of different methods. IL methods show superior efficiency compared with RR methods in either offline training or online training. As we adopt the first-order approximation of MAML, we avoid the expensive computation of Hessian matrices. Thus the meta-learning methods are not much slower than na\"ive IL, and the excess time cost is small enough to omit. It is practical for model selection, hyperparameter tuning, and retraining algorithm selection. Moreover, the low time cost of \model in offline training paves the way for collaboration with RR, \textit{e.g.}, periodically retraining the meta-learners once a year.

\section{Related Work}

\subsection{Stock Trend Forecasting}
Profitable quantitative investment usually depends on precise predictions of stock price trends. In recent years, great efforts~\cite{SFM, Li2018StockPP, Zhao2018, HAN2018, Yoo2021, REST, HIST} have been devoted to designing deep-learning approaches to capture intricate finance patterns, while only a few works study online learning for stock data. 
MASSER~\cite{MASSER}, which mainly focuses on stock movement prediction in the offline setting, introduces Bayesian Online Changepoint Detection in online experiments to detect distribution shifts and updates its meta-learner \textit{after} a detection. Such delayed updates can easily lead to inferior performance in online inference. DDG-DA~\cite{DDGDA}, an advanced RR method, copes with distribution shifts by predicting the future data distribution and resampling the training data for a similar distribution. DDG-DA adapts the training data by assigning samples grouped by periods with different weights \textit{before} a distribution shift happens. By contrast, our data adaptation transforms the features and the label of each sample in a fine-grained way.

\subsection{Meta-Learning}
Meta-learning aims to fast adapt to new tasks only with a few training samples, dubbed \textit{support set}, and generalize well on test samples, dubbed \textit{query set}. 
MAML~\cite{MAML} is the most widely adopted to learn how to fine-tune for good performance on query sets. 
Some works~\cite{FTML, MOLe, W&H} extend MAML to online settings on the assumption that the support set and the corresponding query set come from the same context, \textit{i.e.}, following the same distribution. As such, the meta-learner will quickly remember task-specific information and perform well on a similar query set. However, this assumption cannot hold when discrepancies between the two sets are non-negligible. LLF~\cite{LLF} studies MAML in an offline setting, proving that a predictor optimized by MAML can generalize well against concept drifts. However, the query sets are unlabeled in online settings, and one can only retrain the meta-learner after detecting a shift~\cite{OSAKA}. Consequently, the predictions are still susceptible to distribution shifts. Some methods~\cite{SML, ROLAND} combine incremental learning and meta-learning for recommender systems and dynamic graphs but ignore distribution shifts.
SML~\cite{SML} focuses on model adaptation and proposes a transfer network to convert model parameters on incremental data, which is orthogonal to our work.




\section{Conclusion}
In this work, we propose \modelns, a meta-learning approach to incremental learning for stock trend forecasting. We give two key insights to handle distribution shifts. First, we learn to adapt data into a locally stationary distribution in a fine-grained way. Second, we learn to assign the forecast model with initial parameters which can fast adapt to incremental data and still generalize well against distribution shifts. Experiments on real-world datasets demonstrate that \model is generic and efficient, achieving state-of-the-art predictive performance in stock trend forecasting. 

In the future, we will try to combine
our incremental learning algorithm and rolling retraining to avoid catastrophic forgetting issues after a long-period online incremental learning. 
We also believe the idea of the two-fold adaptation can inspire other applications that encounter the challenge of complex distribution shifts.
  
\begin{acks}
This  work is supported by the National Key Research and Development Program of China  (2022YFE0200500), Shanghai Municipal Science and Technology Major Project  (2021SHZDZX0102), and SJTU Global Strategic Partnership Fund (2021 SJTU-HKUST).
\end{acks}

\bibliographystyle{ACM-Reference-Format}
\balance
\bibliography{online.bib}

\clearpage
\appendix


\section{Implementation Details}
In practice, we employ shared transformation parameters for all time steps of the features when the stock data are taken as time series. Formally, let $\mathbf x = \{x_1, x_2, \cdots, x_L\}$ represent the stock feature, where $L$ is the length of time series. Each element $x_j\in \mathbb R^{d}$ is the feature at step $j$, where $d$ is the number of indicators at each time step. The parameters of the $i$-th transformation head is $\mathbf W_i\in \mathbb R^{d\times d}$ and $\mathbf{b}_i \in \mathbb R^{d}$. At each time step $j$, we first calculate the cosine similarity $\hat s_{ij}$ between each pair of $x_j$ and $\mathbf p_i$, where $\mathbf p_i \in \mathbb R^{d}$. Then, we perform softmax operation over the $N$ heads to obtain the confidence score $s_{ij}$. As such, features of $\mathbf x$ at all time steps are fed into the shared feature adaptation layer with different scores. It is practical to avoid gradient vanishing issues in RNNs. Otherwise, the gradients of more previous steps are too small to train unshared adaptation parameters. 

As for the label adaptation layer, we first employ a linear transformation matrix to cast $\mathbf x\in \mathbb{R}^{Ld}$ into a low-dimensional vector $\mathbf v$. The confidence score for head selection is calculated by the cosine similarity between $\mathbf v$ and $\mathbf p^{\prime}_i$, where $\mathbf p^{\prime}_i$ is a specific prototype for label adaptation and has the same dimension with $\mathbf v$.

\section{Hyperparameter Study}\label{sec:hyperparam}

\begin{figure*}[t]
	\centering
  \captionsetup[subfigure]{margin={0pt, 15pt}}
	\subcaptionbox{Task interval\label{fig:hyper:r}}
	{
    \includegraphics[width=.23\linewidth]{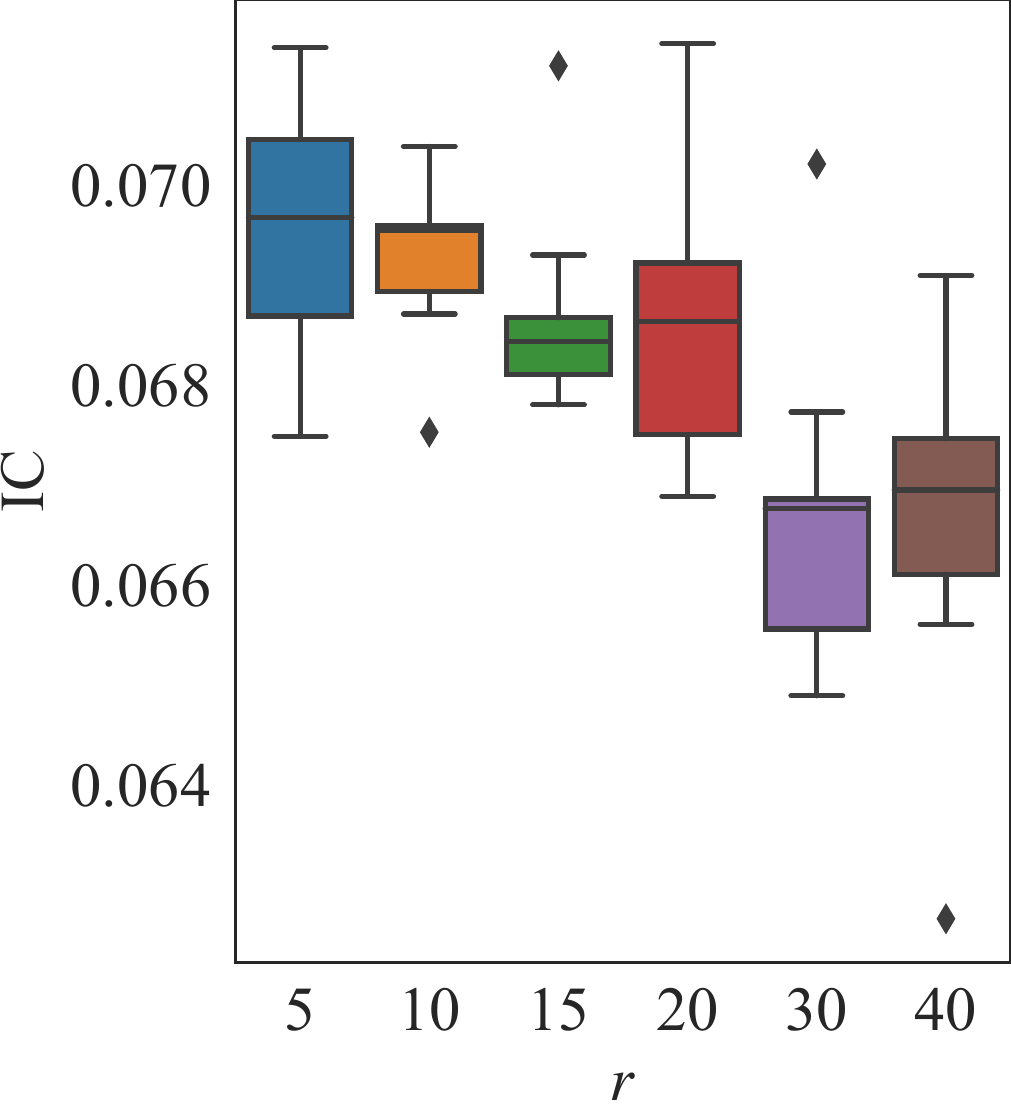}
  }
  \captionsetup[subfigure]{margin={0pt, 15pt}}
	\subcaptionbox{Softmax temparature\label{fig:hyper:tau}}
	{
    \includegraphics[width=.23\linewidth]{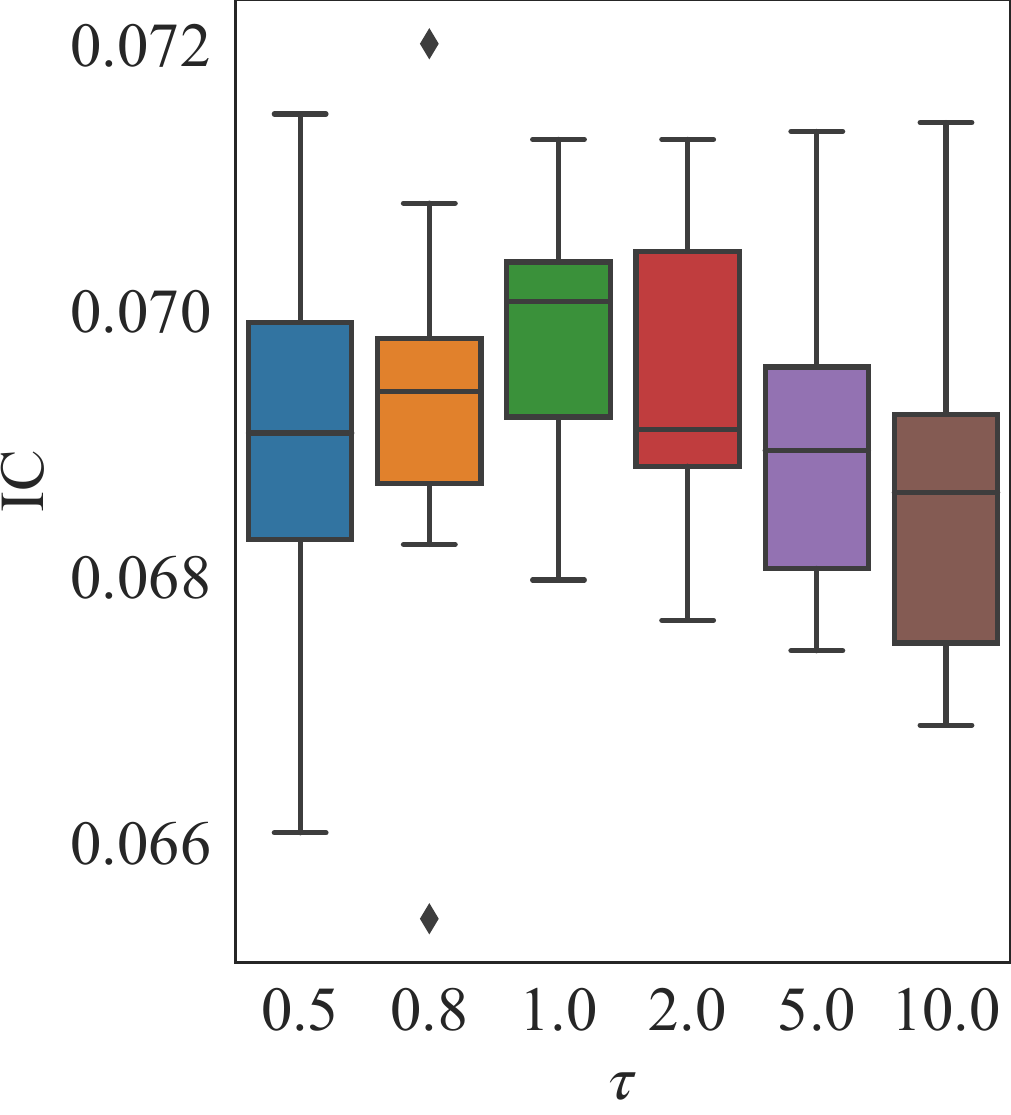}
  }
  \captionsetup[subfigure]{margin={0pt, 15pt}}
	\subcaptionbox{Number of heads\label{fig:hyper:head}}
	{
    \includegraphics[width=.23\linewidth]{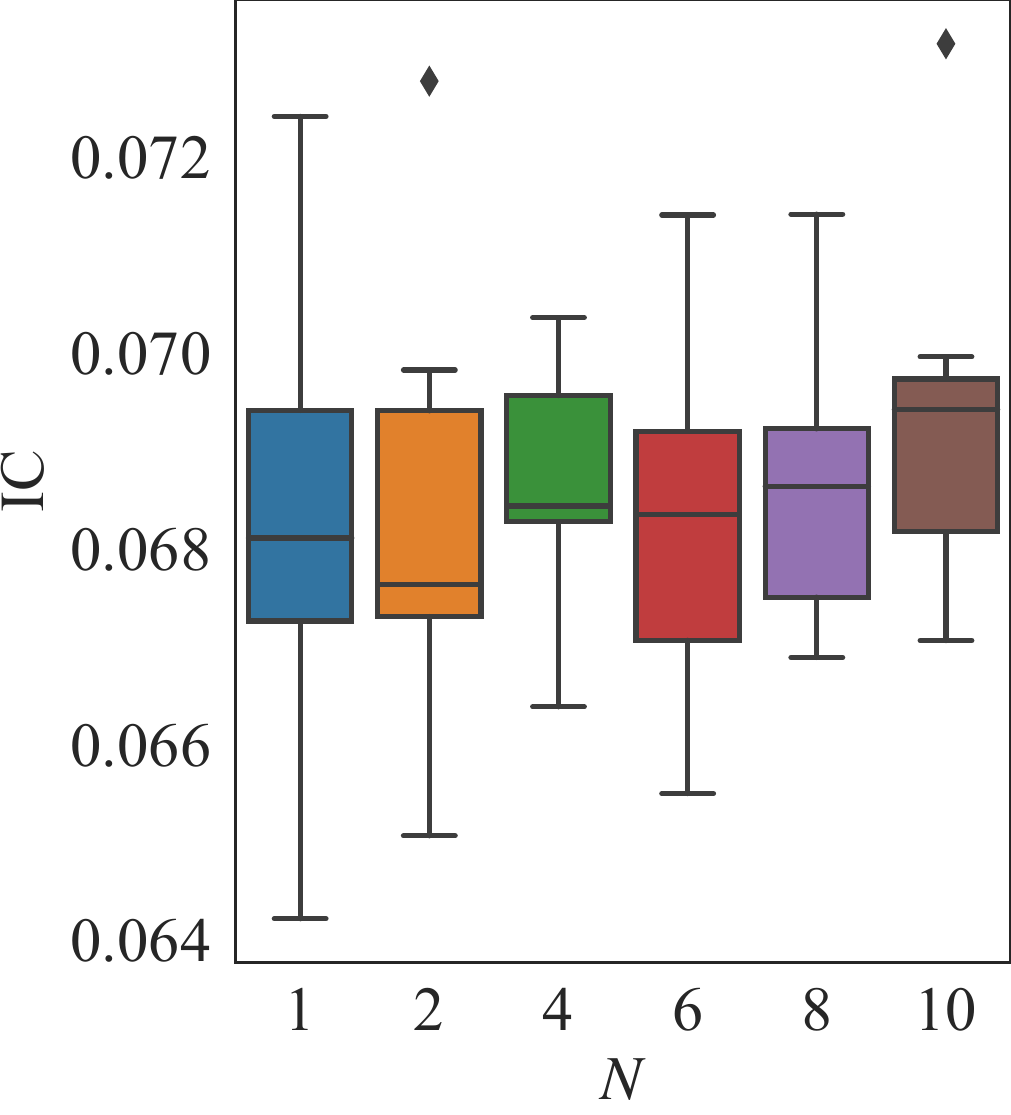}
  }
  \captionsetup[subfigure]{margin={0pt, 15pt}}
	\subcaptionbox{Regularization strength\label{fig:hyper:reg}}
	{
    \includegraphics[width=.23\linewidth]{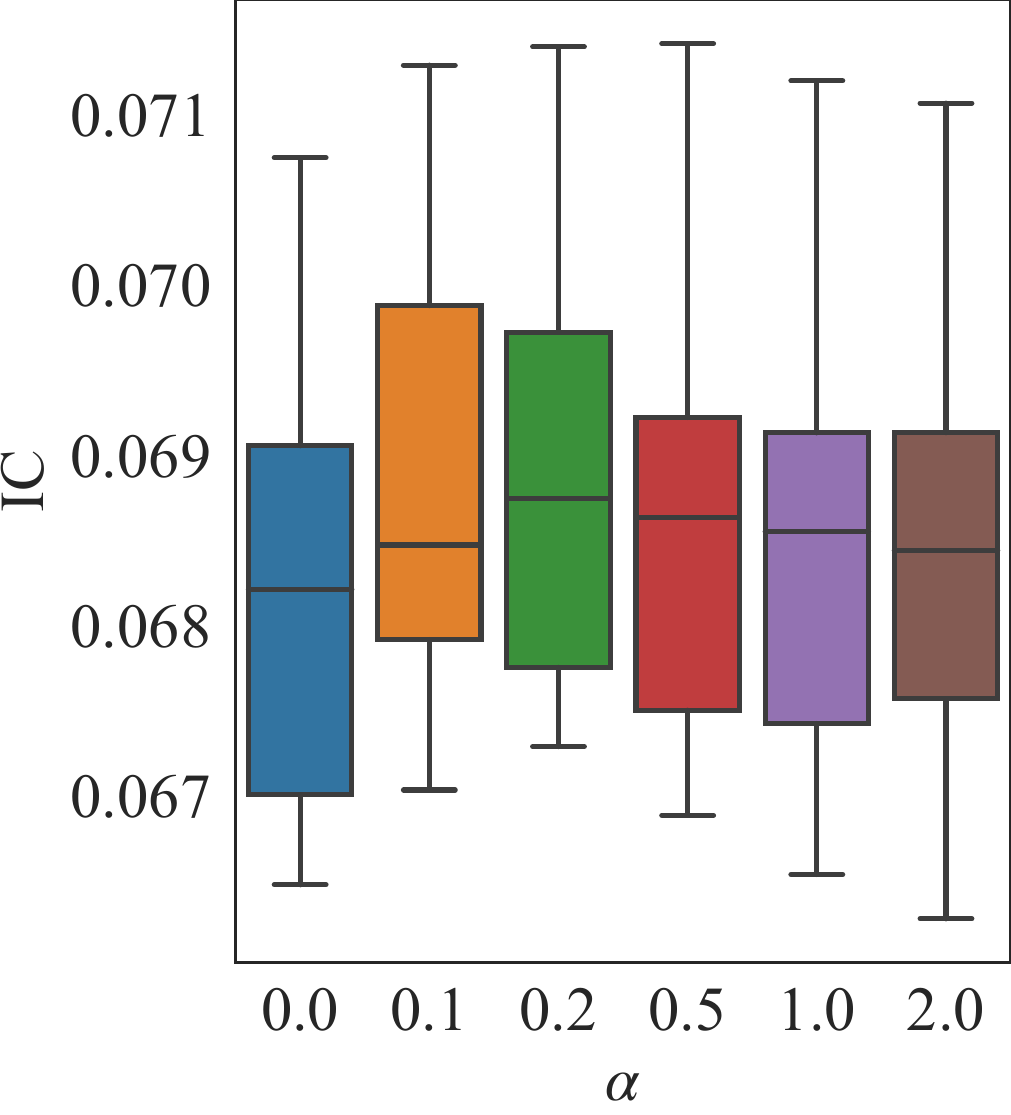}
  }
	\caption{Performance comparison under different hyperparameters on CSI 300. Outliers are marked in diamonds.}
  \label{fig:hyper}
\end{figure*}
In this section, we evaluate the performance of \model on different values of hyperparameters. We only provide the results on CSI 300 with the forecast model implemented by GRU. The results on CSI 300 are similar.

\vspace{0.3em}\noindent\textbf{Task interval $r$. } 
Figure~\ref{fig:hyper:r} shows that more frequent incremental learning with a smaller interval $r$ (\textit{e.g.}, 5 trading days) can lead to better performance. This is because up-to-date patterns are more informative to future trends than previous ones, and it is beneficial to consolidate the forecast model and the meta-learners with new knowledge more timely. It is noteworthy that RR methods with a smaller $r$ will suffer from much more expensive time consumption. In contrast, \model shows superiority in its scalability and is applicable to a small $r$. 

\vspace{0.3em}\noindent\textbf{Softmax temperature $\tau$.}
As shown in Figure~\ref{fig:hyper:tau}, DoubleAdapt achieves the worst performance when we set the softmax temperature $\tau$ to 0.5. 
We reason that the logits from softmax with a small temperature approximate a one-hot vector, \textit{i.e.}, one sample is transformed by merely one head. Thereby, the adaptation layers are trained by fewer samples and can suffer from underfitting issues. Fortunately, it is always safe to set the temperature with a great value and even towards infinity, since a single-head version of \model still achieves outstanding performance in Table~\ref{tab:ablation}. 

\vspace{0.3em}\noindent\textbf{Number of transformation heads $N$.} 
As shown in Figure~\ref{fig:hyper:head}, almost all the multi-head versions of \model outperform the single-head one. Exceptionally, the two-head version shows lower average IC but achieves the highest ceiling value. With a carefully selected temperature (\textit{e.g.}, 1.0), the multi-head version can outperform the single-head one by a larger margin. Note that our incremental learning framework is efficient, and the time cost for hyperparameter tuning is durable. Besides, it is not recommended to use too many heads which can result in more time cost and may also cause overfitting issues. 

\vspace{0.3em}\noindent\textbf{Regularization strength $\alpha$.}
Figure~\ref{fig:hyper:lr} shows that \model without regularization achieves inferior performance, which verifies our proposed regularization technique. Nevertheless, the IC performance gradually decreases with a larger regularization strength which hinders the proposed label adaptation. Thus, it is desirable to set moderate regularization to train \model smoothly. Generally, \model is relatively not sensitive to the regularization strength $\alpha$.

\vspace{0.3em}\noindent\textbf{Online learning rates $\eta_\phi$ and $\eta_\psi$.} As shown in Figure~\ref{fig:hyper:tau}, we evaluate the performance on different learning rates of the two meta-learners during the online training phase, while we use the same pretrained meta-learners before online deployment. 
When we freeze the meta-learners online with both $\eta_\phi$ and $\eta_\psi$ set to zero, \model achieves the worst results. This confirms that it is critical to continually consolidate the meta-learners with new knowledge acquired online. 
Nevertheless, \model still keeps the best performance when we freeze the data adapter with a zero $\eta_\psi$ but fine-tune the model adapter with an appropriate $\eta_\phi$ (0.0005 or 0.001). We conjecture that the pretrained data adapter has learned high-level patterns of distribution shifts, which are shared by the majority of the meta-test set. By contrast, the forecast model should accommodate the up-to-date trends of the stock markets which may not exist in the historical data. This requires the model adapter to be continually fine-tuned on recent incremental data in the online training phase. Otherwise, \model shows inferior performance due to a zero $\eta_\phi$, even if we fine-tune the data adapter with a positive $\eta_\psi$. Besides, as shown in the rightmost column and the bottom row of Figure~\ref{fig:hyper:lr}, the performance degrades with too large learning rates due to catastrophic forgetting issues. It is suggested to tune the hyperparameters of online learning rates on the meta-valid set by using the same pretrained meta-learners for efficiency.

\begin{figure}[t]
    \centering
    \includegraphics[width=0.75\linewidth]{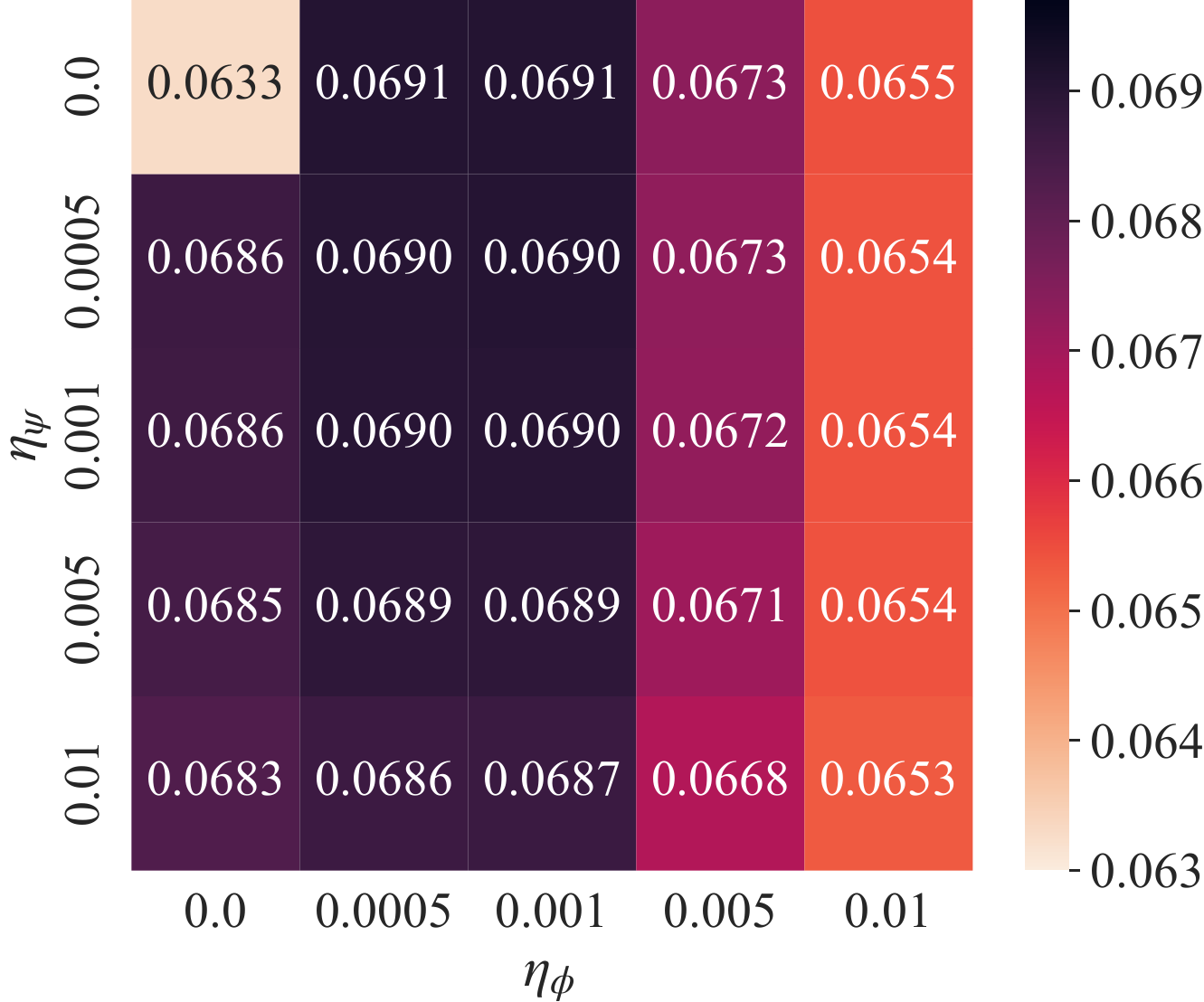}
    \caption{Performance comparison under different online learning rates of the meta-learners on CSI 300. The learning rate $\eta_\phi$ of the model adapter varies on the horizontal axis. The learning rate $\eta_\psi$ of the data adapter varies on the vertical axis.}
    \label{fig:hyper:lr}
\end{figure}

\eat{
\begin{figure*}[ht]
	\centering
  \captionsetup[subfigure]{margin={0pt, 15pt}}
	\subcaptionbox{Task interval\label{fig:hyper:r}}
	{
    \includegraphics[width=.17\linewidth]{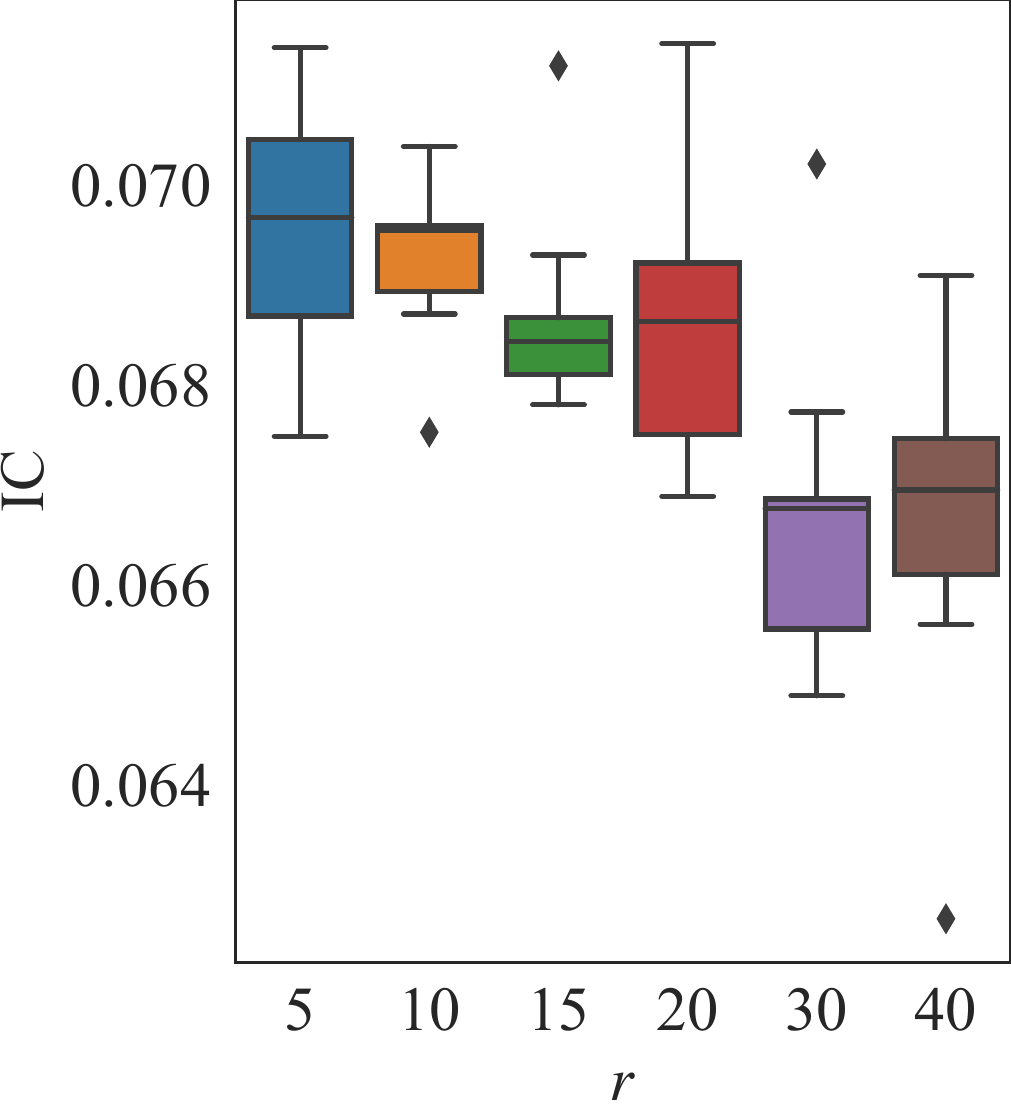}
  }
  \captionsetup[subfigure]{margin={0pt, 15pt}}
	\subcaptionbox{Number of heads\label{fig:hyper:head}}
	{
    \includegraphics[width=.18\linewidth]{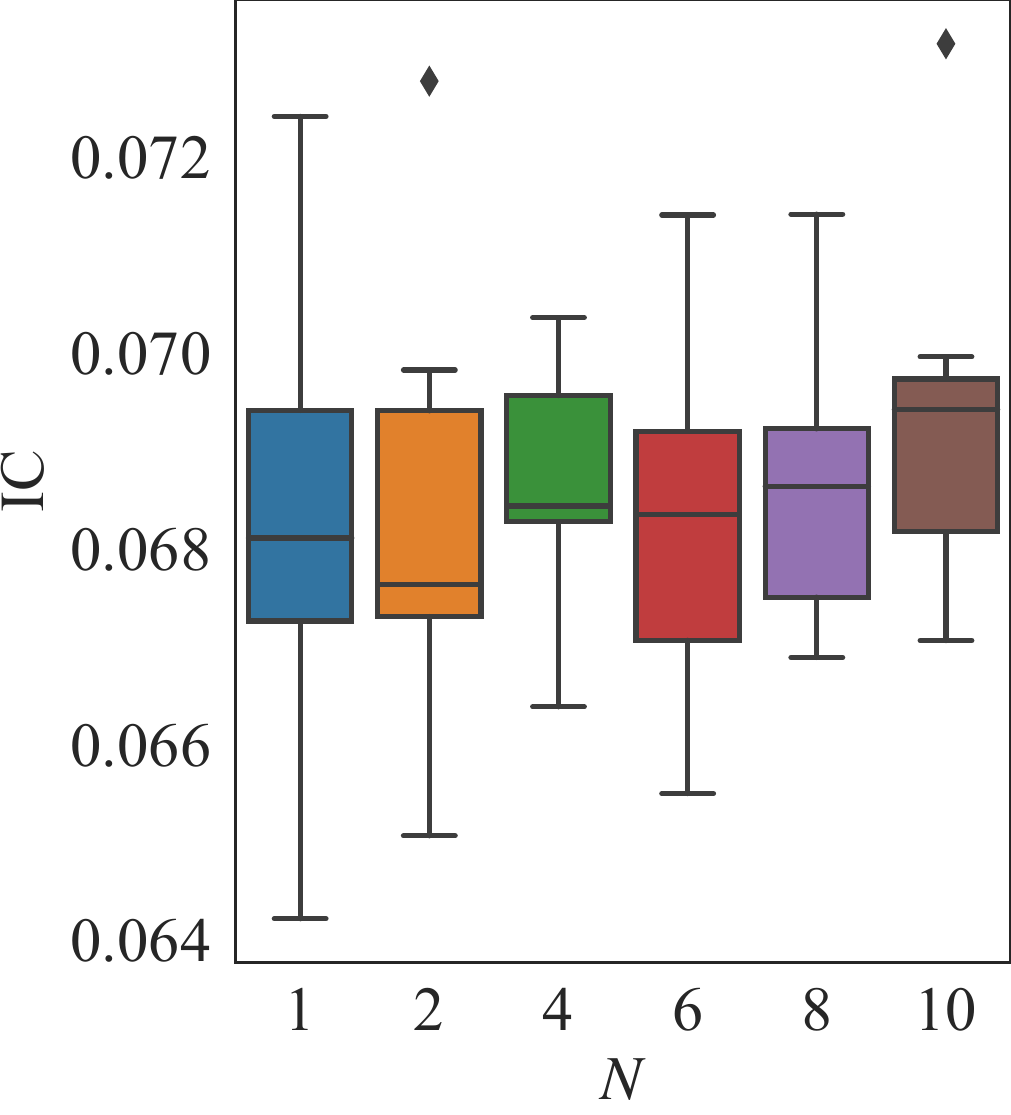}
  }
  \captionsetup[subfigure]{margin={0pt, 15pt}}
	\subcaptionbox{Temparature\label{fig:hyper:tau}}
	{
    \includegraphics[width=.18\linewidth]{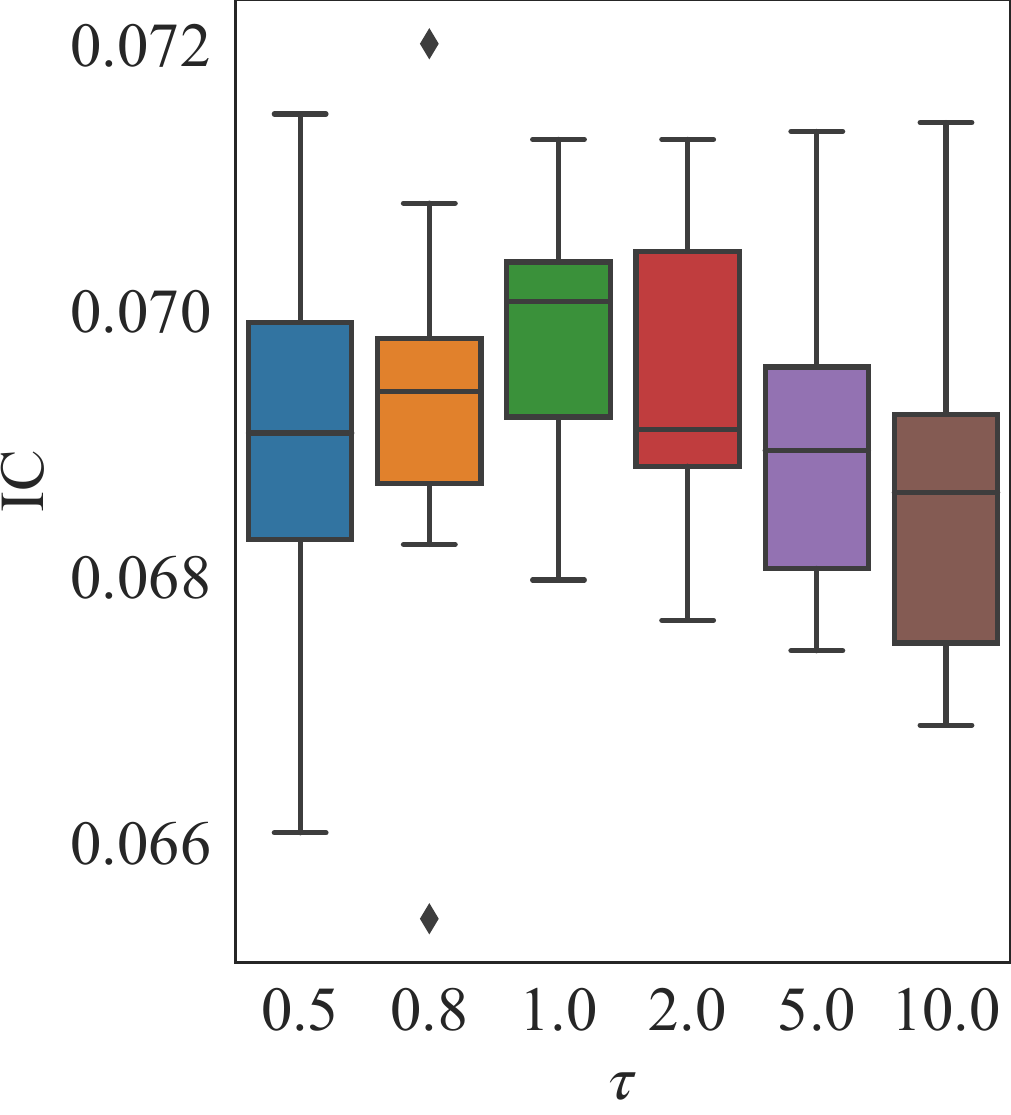}
  }
  \captionsetup[subfigure]{margin={0pt, 15pt}}
	\subcaptionbox{Regularization\label{fig:hyper:reg}}
	{
    \includegraphics[width=.18\linewidth]{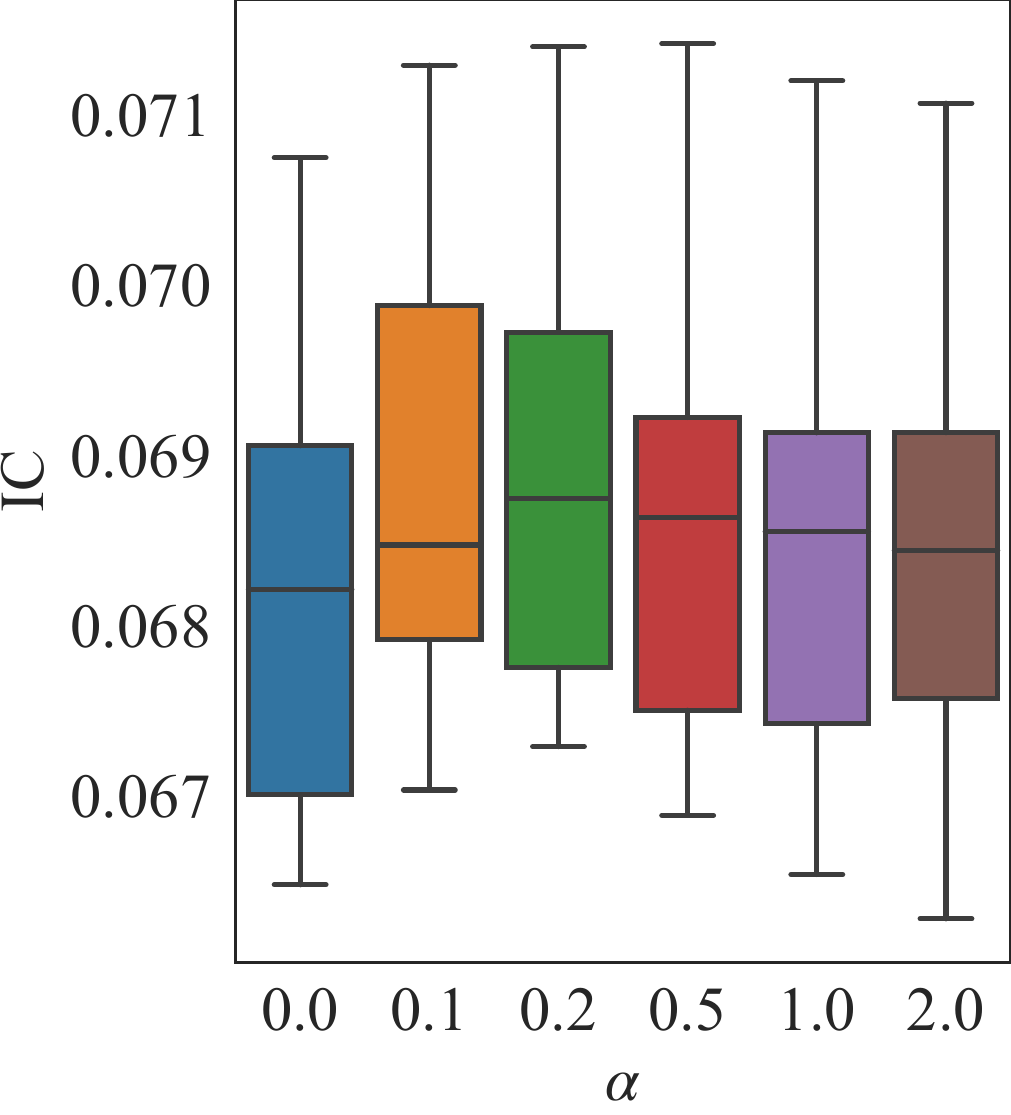}
  }
  \captionsetup[subfigure]{margin={0pt, 0pt}}
	\subcaptionbox{Online learning rates\label{fig:hyper:lr}}
	{
    \includegraphics[width=.18\linewidth]{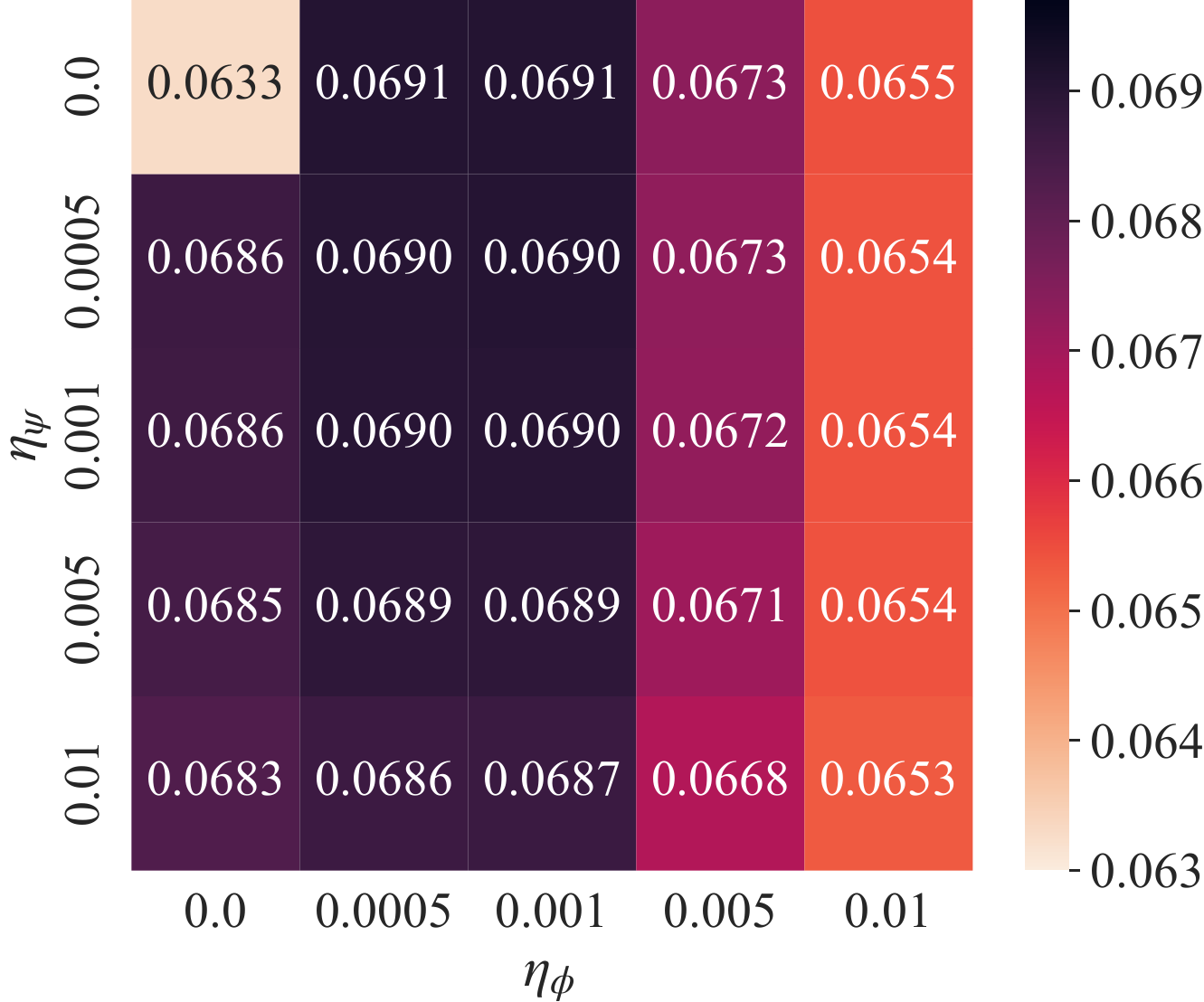}
  }
	\caption{Performance comparison under different hyperparameters on CSI 300. Outliers are marked in diamonds.}
  \label{fig:hyper}
\end{figure*}
}

\section{Approximation of Meta-gradients}
Considering efficiency, we apply first-order approximation in the upper-level optimization of the meta-learners, avoiding the expensive computation of high-order gradients.

To simplify notations, we omit the superscript of the parameters in the following derivations, \textit{i.e.}, $\theta$ for $\theta^{k}$, $\phi$ for $\phi^{k-1}$, and $\psi$ for $\psi^{k-1}$. We further distinguish the parameters of the feature adaptation layer and the label adaptation layer as $\psi_x$ and $\psi_y$, respectively. We also use $\eta$ to represent $\eta_\theta$. 

\begin{figure}[ht]
    \centering
    \includegraphics[width=\linewidth]{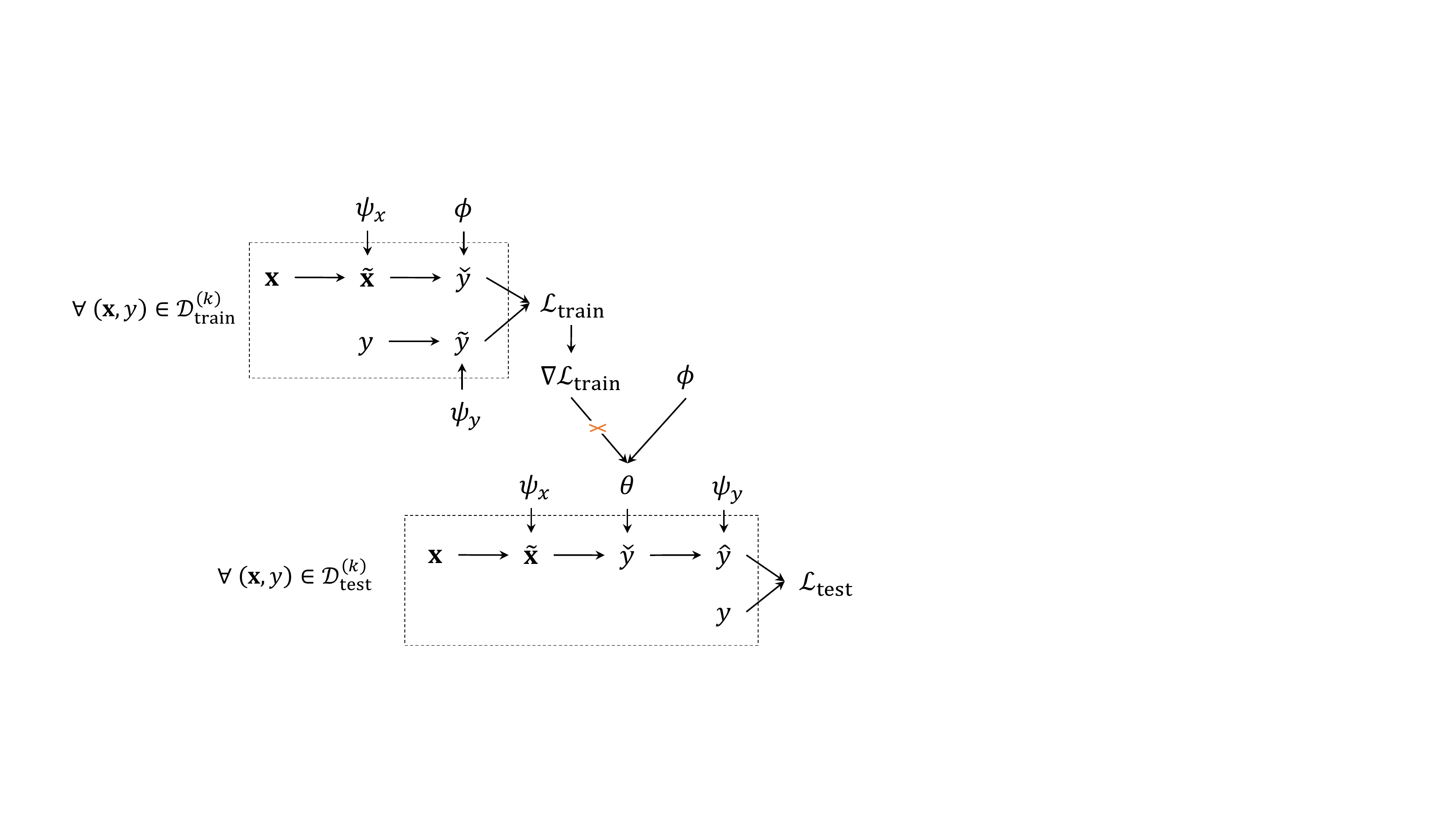}
    \caption{Illustration of our computation graph}
    \label{fig:computation graph}
\end{figure}

\subsection{Gradients of Model Adapter}

The gradients of $\phi$ can be derived by the chain rule:
\begin{equation}
  \begin{aligned}
  \nabla_\phi \mathcal{L}_{\text{test}}(\theta )
  =&\frac{\partial \mathcal{L}_{\text{test}}(\theta )}{\partial \theta }\frac{\partial \theta }{\partial \phi}\\
  =&\frac{\partial \mathcal{L}_{\text{test}}(\theta )}{\partial \theta }\frac{\partial(\phi -\eta \nabla_\phi \mathcal{L}_{\text{train}}(\phi ))}{\partial \phi}\\
  =&\frac{\partial \mathcal{L}_{\text{test}}(\theta )}{\partial \theta }\left(1-\eta\frac{\partial }{\partial \phi}\nabla_\phi \mathcal{L}_{\text{train}}(\phi )\right)\\
  \end{aligned}
\end{equation}

Following MAML~\cite{MAML}, we assume that the second-order gradient is small enough to omit. Thus, we can derive the first-order approximation by
\begin{equation}
\nabla_\phi \mathcal{L}_{\text{test}}(\theta)\approx \frac{\partial \mathcal{L}_{\text{test}}(\theta)}{\partial \theta}=\nabla_\theta \mathcal{L}_{\text{test}}(\theta).
\end{equation}

\subsection{Gradients of Data Adapter}\label{apx:approximation}

As shown in Figure~\ref{fig:computation graph}, the adaptation on features and predictions in the test data involves first-order gradients which directly optimize $\psi_x$ and $\psi_y$. Inspired by~\cite{MPL}, we can also estimate the second-order gradient of $\psi_y$ introduced by $\mathcal L_{\text{train}}$. As we now focus on the second-order gradient, we leave out the first-order one in $\nabla_{\psi_y} \mathcal{L}_{\text{test}}(\theta )$ in the following derivations. 

First, we refomulate the task-specific parameter $\theta$ (\textit{i.e.}, $\theta^{k}$) by
\begin{equation}
\theta = \phi - \eta \mathbb E_{(\tilde{\mathbf x},\tilde y)\in \mathcal{\widetilde D}_{\text{train}} }\nabla_\phi \mathcal L_{\text{MSE}}(F(\mathbf{\tilde x}), \tilde y).
\end{equation}
As $\phi$ is independent of $\psi_y$, we derive the second-order gradients of $\psi_y$ by

\begin{equation}\label{eq:psi_y gradient}
  \begin{aligned}
  \nabla_{\psi_y} \mathcal{L}_{\text{test}}(\theta )
  =&\frac{\partial \mathcal{L}_{\text{test}}(\theta )}{\partial \theta }\frac{\partial \theta }{\partial \psi_y}\\
  =&\frac{\partial \mathcal{L}_{\text{test}}(\theta )}{\partial \theta }\frac{\partial(\phi -\eta \mathbb E_{(\tilde{\mathbf x},\tilde y)\in \mathcal{\widetilde D}_{\text{train}} }\nabla_\phi \mathcal L_{\text{MSE}}(F(\mathbf{\tilde x}), \tilde y))}{\partial {\psi_y}}\\
  =&-\eta\frac{\partial \mathcal{L}_{\text{test}}(\theta )}{\partial \theta }\left(\frac{\partial }{\partial \psi_y}\mathbb E_{(\tilde{\mathbf x},\tilde y)\in \mathcal{\widetilde D}_{\text{train}} }\nabla_\phi \mathcal L_{\text{MSE}}(F(\mathbf{\tilde x}), \tilde y)\right)\\
  \end{aligned}
\end{equation}
Note that $\mathbb E_{(\tilde{\mathbf x},\tilde y)\in \mathcal{\widetilde D}_{\text{train}} }$ is actually a Monte Carlo approximation of $\mathbb E_{(\tilde{\mathbf x},\tilde y)\sim \mathcal{P}_{\text{agent}}(\mathbf x, y) }$. We have
\begin{equation} \label{eq:second-order gradient}
\begin{aligned}
  &\frac{\partial }{\partial \psi_y}\mathbb E_{(\tilde{\mathbf x},\tilde y)\in \mathcal{\widetilde D}_{\text{train}} }\nabla_\phi \mathcal L_{\text{MSE}}(F(\mathbf{\tilde x}), \tilde y)\\
  =&\frac{\partial }{\partial \psi_y}\mathbb E_{(\tilde{\mathbf x},\tilde y)\sim \mathcal{P}_{\text{agent}}(\mathbf x, y) }\nabla_\phi \mathcal L_{\text{MSE}}(F(\mathbf{\tilde x}), \tilde y)\\
  =&\mathbb E_{\tilde{\mathbf x}\sim\mathcal P_{\text{agent}}(\mathbf x)}\frac{\partial }{\partial \psi_y}\mathbb E_{\tilde y\sim \mathcal P_{\text{agent}}(y|\mathbf x)}\nabla_\phi \mathcal L_{\text{MSE}}(F(\mathbf{\tilde x}), \tilde y)),
  \end{aligned}
\end{equation}
where we extract the expectation on $\tilde{\mathbf x}$ out of the gradient on $\psi_y$ as $\tilde{\mathbf x}$ is independent of $\psi_y$.

Since $\nabla_\phi\mathcal L_{\text{MSE}}(F(\mathbf{\tilde x}), \tilde y))$ has no dependency on $\psi_y$, except for via $\tilde y$, we can apply the REINFORCE rule~\cite{REINFORCE} to achieve
\begin{equation} \label{eq:REINFORCE}
\begin{aligned}
&\frac{\partial }{\partial \psi_y}\mathbb E_{\tilde y\sim \mathcal P_{\text{agent}}(y|\mathbf x)}\left[\nabla_\phi \mathcal L_{\text{MSE}}(F(\mathbf{\tilde x}), \tilde y))\right]\\
=&\mathbb E_{\tilde y\sim \mathcal P_{\text{agent}}(y|\mathbf x)}\left[\nabla_\phi \mathcal L_{\text{MSE}}(F(\mathbf{\tilde x}), \tilde y)\cdot \frac{\partial \log P(\tilde y\mid\mathbf x,y;\psi_y)}{\partial \psi_y}\right],
\end{aligned}
\end{equation}
where $P(\tilde y\mid\mathbf x,y;\psi_y)$ is the probability density of $\mathcal P_{\text{agent}}(y|\mathbf x)$. 

Assume the conditional probability follows a normal distribution, \textit{i.e.}, $P(\tilde y \mid \mathbf x,y) = \mathcal N(y,\sigma^2)$, where $\sigma$ is a constant. We have
\begin{equation} \label{eq:logMSE}
\begin{aligned}
\frac{\partial \log P(\tilde y|\mathbf x,y;\psi_y)}{\partial \psi_y}
=&\frac{\partial \left(-\frac{1}{2} \log \left(2 \pi \sigma^2\right)-\frac{1}{2 \sigma^2}\left(\tilde y-y\right)^2\right)}{\partial \psi_y}\\
=&\frac{\partial \left(-\frac{1}{2} \log \left(2 \pi \sigma^2\right)-\frac{1}{2 \sigma^2}\mathcal{L}_{\text{MSE}}(\tilde y, y)\right)}{\partial \psi_y}\\
=&-\frac{1}{2\sigma^2}\frac{\partial \mathcal{L}_{\text{MSE}}(\tilde y, y)}{\partial \psi_y}
\end{aligned}
\end{equation}
Then Eq.~(\ref{eq:second-order gradient}) becomes
\begin{equation} \label{eq:flatten second-order gradient} 
\begin{aligned}
&\frac{\partial }{\partial \psi_y}\mathbb E_{(\tilde{\mathbf x},\tilde y)\in \mathcal{\widetilde D}_{\text{train}} }\nabla_\phi \mathcal L_{\text{MSE}}(F(\mathbf{\tilde x}), \tilde y)\\
=&\frac{-1}{2\sigma^2}\mathbb E_{\tilde{\mathbf x}\sim\mathcal P_{\text{agent}}(\mathbf x)}\mathbb E_{\tilde y\sim \mathcal P_{\text{agent}}(y|\mathbf x)}\left[\nabla_\phi \mathcal L_{\text{MSE}}(F(\mathbf{\tilde x}), \tilde y))\frac{\partial \mathcal{L}_{\text{MSE}}(\tilde y, y)}{\partial \psi_y}\right]\\
=&\frac{-1}{2\sigma^2}\mathbb E_{(\tilde{\mathbf x},\tilde y)\in \mathcal{\widetilde D}_{\text{train}} }\left[\nabla_\phi \mathcal L_{\text{MSE}}(F(\mathbf{\tilde x}), \tilde y))\frac{\partial \mathcal{L}_{\text{MSE}}(\tilde y, y)}{\partial \psi_y}\right]\\
=&\frac{-1}{2\sigma^2}\nabla_\phi \mathcal L_{\text{train}}(\phi)\frac{\partial\ \mathbb E_{(\tilde{\mathbf x},\tilde y)\in \mathcal{\widetilde D}_{\text{train}} }\mathcal{L}_{\text{MSE}}(\tilde y, y)}{\partial \psi_y},\\
=&\frac{-1}{2\sigma^2}\nabla_\phi \mathcal L_{\text{train}}(\phi)\frac{\partial\mathcal{L}_{\text{reg}}}{\partial \psi_y},\\
\end{aligned}
\end{equation}
where $\mathcal{L}_{\text{MSE}}(\tilde y, y)$ is only calculated on the training samples, and the expectation is actually the regularization loss $\mathcal L_{\text{reg}}$ in our paper. Then, we substitute Eq.~(\ref{eq:flatten second-order gradient}) into Eq.~(\ref{eq:psi_y gradient}) to obtain
\begin{equation}
\begin{aligned}
    \nabla_{\psi_y} \mathcal{L}_{\text{test}}(\theta )
  =\frac{\eta}{2\sigma^2}[\nabla_\theta\mathcal{L}_{\text{test}}(\theta)]^\top\cdot\nabla_\phi \mathcal L_{\text{train}}(\phi)\cdot \frac{\partial \mathcal{L}_{\text{reg}}}{\partial \psi_y}.
\end{aligned}
\end{equation}
With Taylor expansion, we have
\begin{equation} 
\mathcal L_{\text{test}}(\phi)\approx\mathcal L_{\text{test}}(\theta)+[\nabla_\theta\mathcal{L}_{\text{test}}(\theta )]^\top(\phi-\theta).
\end{equation}
Note that
\begin{equation} 
\theta=\phi-\eta\nabla_\phi\mathcal L_{\text{train}}(\phi).
\end{equation}
Thus, 
\begin{equation} 
\mathcal L_{\text{test}}(\phi)-\mathcal L_{\text{test}}(\theta)\approx\eta[\nabla_\theta\mathcal{L}_{\text{test}}(\theta)]^\top\nabla_\phi \mathcal L_{\text{train}}(\phi)
\end{equation}
Finally, the second-order gradient of $\psi_y$ is approximated by
\begin{equation} 
\nabla_{\psi_y} \mathcal{L}_{\text{test}}(\theta )=\frac{\mathcal L_{\text{test}}(\phi)-\mathcal L_{\text{test}}(\theta)}{2\sigma^2}\cdot\frac{\partial \mathcal{L}_{\text{reg}}}{\partial \psi_y},
\end{equation}
where we set $\sigma$ as a hyperparameter. Thus the coefficient of gradients by $\mathcal L_{\text{reg}}$ is similar to $\alpha$ which controls the regularization strength when $\mathcal L_{\text{test}}(\phi)\ge\mathcal L_{\text{test}}(\theta)$. If otherwise, the updated parameters result in a greater loss, meaning that the incremental data is noisy and perhaps we should make the adapted labels $\tilde y$ more different from the original $y$.

\end{document}